\documentclass[a4paper,10pt,papersize]{article}

\RequirePackage[T1]{fontenc}
\RequirePackage{fix-cm}
\usepackage{amsmath,amsfonts,amsthm}
\usepackage{graphicx} 
\usepackage[unicode,final]{hyperref} 
\usepackage{cite}
\usepackage{caption}
\usepackage{mathtools}
\usepackage{newtxtext}
\usepackage[varvw]{newtxmath}
\usepackage{bm}
\usepackage{textgreek}
\usepackage[T1]{fontenc}
\usepackage[utf8]{inputenc}
\usepackage{color}

\allowdisplaybreaks[4]
\makeatletter

\renewcommand{\proofname}{Proof}
\makeatother
\usepackage{soul}

\usepackage{booktabs}
\usepackage{mathtools}
\usepackage{textcomp,marvosym}

\begin{document}
\def\rem#1{{{}}}
\newcommand{\changed}[1]{{#1}}
\newcommand{\delete}[1]{{{\!\!}}}
\title{Data-based stochastic modeling reveals sources of activity bursts in single-cell TGF-β signaling}
%
%
%
%
\author{
  Niklas~Kolbe\textsuperscript{1\footnote{Corresponding author}~\,\Yinyang} \and
  Lorenz~Hexemer\textsuperscript{2\Yinyang} \and
  Lukas-Malte~Bammert\textsuperscript{3} \and
  Alexander~Loewer\textsuperscript{4} \and
  M\'aria~Luk\'a\v{c}ov\'a-Medviďov\'a\textsuperscript{3\ddag} \and
  Stefan~Legewie\textsuperscript{5*\ddag}
 }
%
\date{
  \small
  $^1$Institut für Geometrie und Praktische Mathematik, RWTH Aachen University, Aachen, Germany \smallskip\\
  $^2$Institute of Molecular Biology (IMB), Mainz, Germany \smallskip\\
  $^3$Institute of Mathematics, Johannes Gutenberg-University,  Mainz, Germany \smallskip\\
  $^4$Systems Biology of the Stress Response, Department of Biology, Technische Universität Darmstadt, Darmstadt, Germany \smallskip\\
  $^5$Department of Systems Biology and Stuttgart Research Center for Systems Biology (SRCSB), University of Stuttgart, Stuttgart, Germany \smallskip\\
  {\tt kolbe@igpm.rwth-aachen.de},\ \
  {\tt l.ripka@imb-mainz.de},\ \
  {\tt lbammert@students.uni-mainz.de}, \ \
  {\tt loewer@bio.tu-darmstadt.de}, \ \
  {\tt lukacova@uni-mainz.de}, \ \
  {\tt legewie@iig.uni-stuttgart.de}\\
  \bigskip
  \Yinyang These authors contributed equally to this work.\\
  \ddag These authors also contributed equally to this work.
}
\maketitle

\begin{abstract}
	Cells sense their surrounding by employing intracellular signaling pathways that transmit hormonal signals from the cell membrane to the nucleus. TGF-β/SMAD signaling encodes various cell fates, controls tissue homeostasis and is deregulated in diseases such as cancer. The pathway shows strong heterogeneity at the single-cell level, but quantitative insights into mechanisms underlying fluctuations at various time scales are still missing, partly due to inefficiency in the calibration of stochastic models that mechanistically describe signaling processes. In this work we analyze single-cell TGF-β/SMAD signaling and show that it exhibits temporal stochastic bursts which are dose-dependent and whose number and magnitude correlate with cell migration. We propose a stochastic modeling approach to mechanistically describe these pathway fluctuations with high computational efficiency. Employing high-order numerical integration and fitting to burst statistics we enable efficient quantitative parameter estimation and discriminate models that assume noise in different reactions at the receptor level. This modeling approach suggests that stochasticity in the internalization of TGF-β receptors into endosomes plays a key role in the observed temporal bursting. Further, the model predicts the single-cell dynamics of TGF-β/SMAD signaling in untested conditions, e.g., successfully reflects memory effects of signaling noise and cellular sensitivity towards repeated stimulation. Taken together, our computational framework based on burst analysis, noise modeling and path computation scheme is a suitable tool for the data-based modeling of complex signaling pathways, capable of identifying the source of temporal noise. 
\end{abstract}

\section*{Introduction}

During development and homeostasis of mammalian tissues,  proliferation and differentiation are coordinated among thousands of cells. To communicate with each other, cells use cytokines, a group of extracellular signaling molecules. The cytokine transforming growth factor beta (TGF-β) and other members of the TGF-β superfamily play an important role in tissue homeostasis, as they induce antiproliferative and apoptotic responses in adult cells to effectively limit tissue growth~\cite{siegel2003, Heldin2009}.
Furthermore TGF-β signaling induces a loss of cell-cell junctions, cytoskeletal reorganization and migration~\cite{Moustakas2007}, thereby allowing epithelial cells to evade from their original location by acquiring a migratory, mesenchymal phenotype in a process called epithelial-to-mesenchymal transition (EMT).

Since the TGF-β signaling pathway is involved in such central regulatory processes, its deregulation is associated with diseases like fibrosis and cancer. In cancer progression, TGF-β signaling plays a dual role:
In normal cells the pathway typically provides a cytostatic response and thus acts as a tumour suppressor. In early tumor stages, the tumor-suppressive function is frequently lost, whereas in late-stage tumors TGF-β signaling may induce EMT and promote metastasis~\cite{Ikushima2010}. Hence, TGF-β signaling undergoes a specificity switch from a tumor-suppressing to a migration-enhancing function. To better understand this specificity switch, deeper insights into cellular information processing in the TGF-β pathway are required.

The molecular mechanisms of TGF-β signal transmission within cells are well characterized: Signaling is initiated by binding of TGF-β to the TGF-β receptor type II (TGFBR2). This receptor then recruits a type I receptor (TGFBR1) to form the receptor-ligand complex, which phosphorylates SMAD2 and SMAD3, 
the intracellular transducers of the signal. In the cytoplasm, phosphorylated SMAD2/3 proteins form trimers with another protein called SMAD4, and the SMAD trimers translocate into the nucleus to act as transcription factors regulating the expression of downstream target genes. Among the induced genes are both positive and negative feedback regulators, e.g., SMAD7 which inhibits TGF-β receptors by inhibiting their catalytic activity and inducing their degradation~\cite{Schmierer2007}. 

Previous studies indicate that the cellular response to TGF-β stimulation is correlated to the temporal dynamics of SMAD activation~\cite{Nicolas2003, Strasene.2018, sorreEncodingTemporalSignals2014}. For instance, in cancer cell lines EMT and cell migration seems to be induced upon transient SMAD activation, while the cytostatic response requires sustained signaling~\cite{Nicolas2003}. Moreover, we recently analyzed the dynamics of SMAD nuclear translocation in single MCF10A cells using live-cell imaging and found that cells with sustained signaling tended to divide less when compared to cells showing a transient signal~\cite{Strasene.2018}. Due to high temporal and spatial resolution, this data allowed us to accurately observe the temporal heterogeneity of SMAD signaling at the single-cell level. Interestingly, we found temporal fluctuations in the nuclear translocation of SMAD2 on an intermediate time scale of 30 min to 2 hours which we refer to as \emph{activity bursts} in the following. These bursts may be physiologically relevant, as stochastic or periodic pulsatile changes in signaling proteins were found to have an influence on the cellular response for several other pathways including calcium~\cite{Cai2008}, NF-kB~\cite{Ashall2009,Sakai2017}, ERK~\cite{Aoki2013,Albeck2013} and P53~\cite{Purvis2012} signaling.  

The temporal regulation of SMAD translocation is well studied and understood by deterministic mathematical models describing the average dynamics of large cell populations, see e.g.~\cite{Strasene.2018, paulsenNegativeFeedbackBone2011, Vilar2006, Wegner2012, Zi2007}. Using these models, ligand degradation due to receptor internalization and the negative feedback via SMAD7 have been identified as important factors in the termination of signal. However, deterministic models do not account for the temporal activity bursts we observe at the single-cell level. For a better understanding of the pathway dynamics and to reproduce the effect of bursting events on the pathway, stochastic simulations are required. A common but computationally expensive way to integrate stochastic effects in time trajectories is to use the Gillespie algorithm~\cite{Gillespie1976}, also known as kinetic Monte Carlo method. The algorithm \delete{does not deliver a formal} \changed{delivers a} solution of the underlying chemical master equation \delete{but exact results can be achieved} by sampling from a large number of realizations~\cite{Gillespie2007, ElSamad2005}. \changed{Equivalently, the chemical Langevin equation introduces a system of stochastic differential equations (SDEs), that assumes noise in the concentrations of the chemical species to approximate the stochastic dynamics}\rem{R2MA2}~\cite{kurtzStrongApproximationTheorems1978, gillespieChemicalLangevinEquation2000}. Other types of SDE \changed{and random ordinary differential equation} models have been applied to cellular dynamics for example in~\cite{samoilovStochasticAmplificationSignaling2005, suelExcitableGeneRegulatory2006, yiEnhancementInternalnoiseCoherence2006}\rem{R2MA3}. The complexity of the \changed{TGF-β}\rem{R2MA4} system under investigation is reflected by a large number of dynamic variables and parameters acting on different time scales leading to a stiff system of ordinary differential equations (ODEs). This reduces the efficiency of stochastic simulations and complicates the quantitative calibration of stochastic models based on large-scale single-cell datasets.  

\changed{Species that are present in low copy numbers contribute most to the stochasticity. In contrast, species with large copy numbers or fast reaction rates necessitate smaller time steps in the Gillespie algorithm, thereby increasing the required computing time. In common splitting methods, either the species ~\cite{rao2003} or the reactions are split into a slow and a fast group~\cite{salis2005, weinan2007}, or combinations of both~\cite{cao2005}  under quasi-steady-state or partial equilibrium assumption. In other hybrid methods, based on a separation of time scales, fast components are modeled deterministically while other components are treated with a stochastic model~\cite{Kiehl2004}. Further hybrid methods determine their stochastic species based on the variance \cite{lotstedt2006, hellander2007}. Derived from fundamental chemical reactions of simple molecules without conformational changes, all of these methods assume Poisson- or normal distributions of reaction events.}\rem{R2A1}

 \delete{To circumvent this problem, we propose a hybrid deterministic-stochastic modeling framework, in which stochasticity by means of SDEs is only applied to a subset of suitable, selected variables within the system to improve the efficiency of simulations.} 
 
 \changed{In contrast to these methods we propose a hybrid deterministic-stochastic modeling framework, which allows for multi-parametric stochastic processes corresponding to more complex processes, still isolating the underlying and unknown complexity in a systems biology manner. To this end our efficient approach applies stochasticity by means of SDEs to a subset of selected variables within the system. We further propose an analysis framework that uses species splitting to our advantage since it offers the possibility to attribute sources of noise.}\rem{R2A1} 
 To quantitatively reproduce bursting events in the SMAD signaling pathway, we apply an automated detection algorithm enabling us to analyze how burst properties vary across the population. Applying the same burst analysis to both the measured data and the simulations, we define an objective function and thereby compare different model variants that consider stochasticity in distinct parts of the signaling cascade to the data. Thus, we were able to determine the contribution of different pathway reactions to generating activity bursts in SMAD signaling and found that TGF-β receptor internalization may play a key role in this respect. Finally, we compared the predictions of our model to experimental data in untested conditions and show that our model faithfully reproduces the temporal heterogeneity of the SMAD signaling network. The proposed framework may be generally applicable to quantitatively model stochastic biochemical signaling networks based on live-cell imaging data.



	\section*{Results}
	\begin{figure}
	\caption{Quantification of bursts in TGF-β-induced SMAD translocation}\label{fig:data}
		\includegraphics[width=\linewidth]{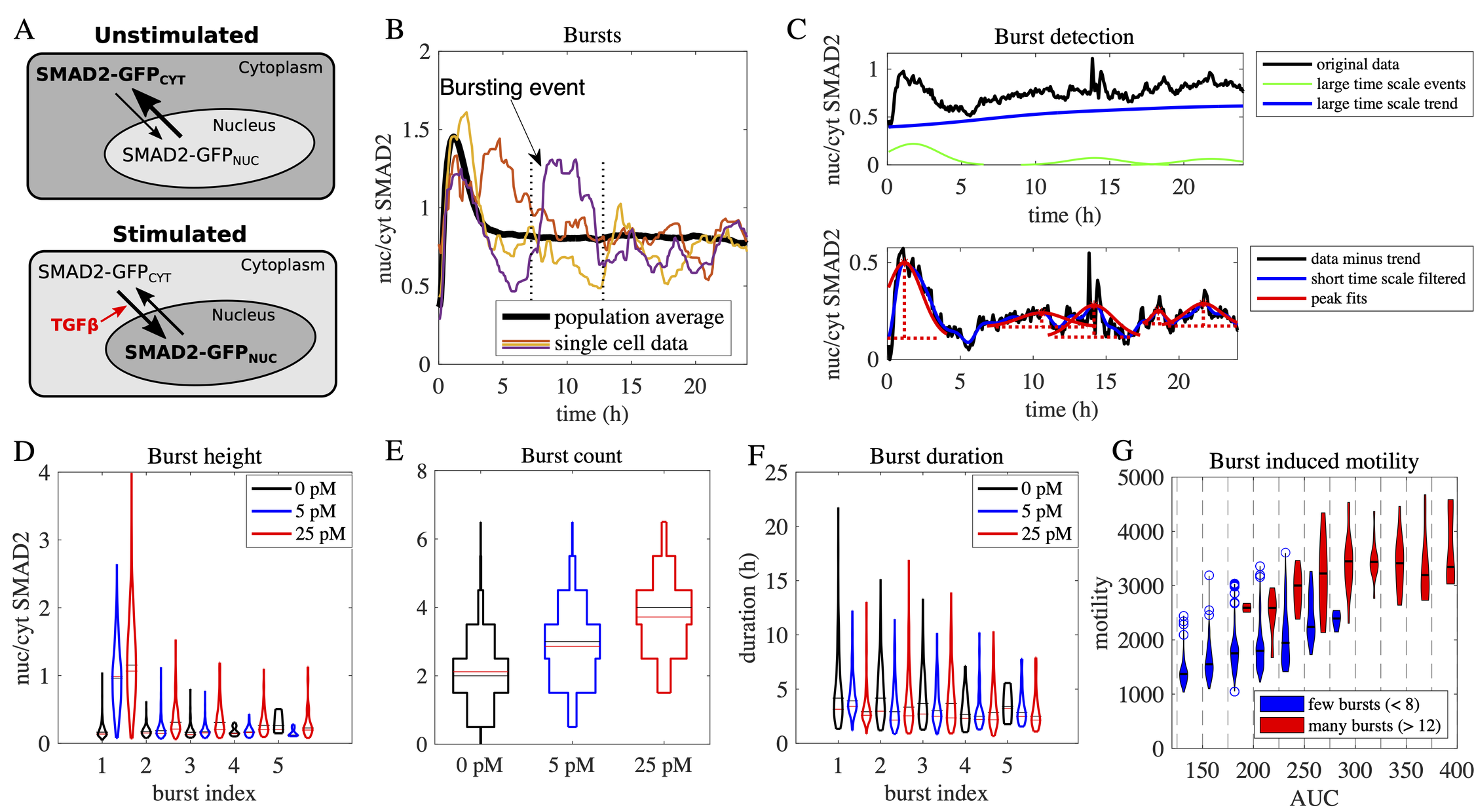}
		
		\begin{flushleft}A:\ Scheme of microscopy-based single-cell analysis of TGF-β signaling.
                  In unstimulated cells the SMAD2-GFP fusion protein is primarily located in the cytoplasm (top). TGF-β stimulation promotes the transport of SMAD2-GFP to the nucleus and hence increases the nuc/cyt SMAD2 ratio that was quantified from microscopy images (bottom). See~\cite{Strasene.2018} for details on the data analysis.
			
			B:\ Bursts in SMAD2 nuclear translocation. Time-resolved measurements of the nuc/cyt SMAD2 ratio of four
			individual cells (colored lines) are compared to the population average response over 352 cells (black line) upon stimulation with 100 pM TGF-β. A stochastic deflection (burst) is highlighted (arrow, dotted lines indicate beginning and end).
			
			C:\ Burst detection process. Bursts are detected in the nuc/cyt SMAD2 ratio of a single-cell upon stimulation with TGF-β. Top: Peaks in the measured trajectory of the nuc/cyt SMAD2 ratio (black line) are approximated by a sum of Gaussian hills (green lines) which are subtracted from a smoothened version of the signal to estimate the long-term trend (blue line). Bottom: Afterwards, the long term trend is subtracted from the measurement data (black line), short-term effects are smoothed out by a Gaussian filter (blue line) and the result is analyzed by a peak finding algorithm which detects 5 bursts in this example (indicated in red). \changed{As pathway events might occur shortly after each other we allowed for overlapping bursts.}\rem{R1A1} The bursts are characterized by their height and duration (red dashed lines). See Methods for details.
			
			D:\ Burst height increases with TGF-β dose. Violin plots show the distributions of the burst height for the first 5 bursts (burst index) in the nuc/cyt SMAD2 ratio trajectories upon stimulation with 5 pM (blue) and 25 pM (red) TGF-β and in a control population (0~pM). Median and mean are marked by red and black horizontal lines.
			
			E:\ Burst count increases with TGF-β dose. Distribution of burst counts in the nuc/cyt SMAD2 ratio trajectories after stimulation with 5~pM and 25~pM TGF-β and in a control population (0~pM). Higher doses of TGF-β lead to higher burst counts. Median and mean are marked by red and black horizontal lines.

            F:\ Burst duration is independent of TGF-β dose. Violin plots show the distributions of burst duration for the first 5 bursts (burst index) in the nuc/cyt SMAD ratio trajectories upon stimulation with 5 pM and 25 pM TGF-β and in a control population (0~pM). Median and mean are marked by red and black horizontal lines.
            
            G:\ Bursts count correlates to cell motility at the single-cell level. Single cells were binned according to their area-under-curve (AUC), measured as the integral of the nuc/cyt SMAD2 ratio over all time points (x-axis), and were further sub-classified in terms of their total burst count (see legend). The cell motility (y-axis) was calculated by taking the sum over the distance moved by the cells in consecutive time points. Among cells with similar total signal (AUC), cells with 8 or less bursts (blue) exhibit lower motility than cells with 12 or more bursts (red). 
			
		\end{flushleft}
\end{figure}

\subsection*{Quantification of temporal stochasticity in single cells by burst analysis}

To analyze activity bursts in TGF-β/SMAD signaling, we used previously published measurements of SMAD2 nuclear translocation at the single-cell level~\cite{Strasene.2018}.
In these experiments, MCF10A breast epithelial cells stably expressing SMAD2 fused to a fluorescent protein were analyzed by time-resolved live-cell microscopy. The nuclear and cytoplasmic SMAD2 concentrations were quantified over a 24h time interval after stimulation with different concentrations of TGF-β. To determine pathway activity, the ratio of nuclear to cytoplasmic SMAD2 (nuc/cyt SMAD2 ratio) was used as a quantitative measure of the translocation of SMAD2 to the nucleus (see Fig.~\ref{fig:data}).

\changed{The analysis in~\cite{Strasene.2018} showed that upon}\rem{R1A1} stimulation with a saturating TGF-$\beta$ stimulus (100 pM), the population-average response summarizing 352 single cells shows a transient peak that is followed by a plateau above the baseline  (black line in Fig.~\ref{fig:data}B). 
\changed{In contrast to~\cite{Strasene.2018}, where heterogeneity in the cell population was investigated, we here focus on temporal heterogeneity.}\rem{R1A1}
At the single-cell level, \changed{we observed} \delete{the time courses were highly heterogeneous and} fluctuations in the nuc/cyt SMAD2 ratio \delete{were observed} on three different time scales: On a short time scale of less then 100 minutes, there are temporally uncorrelated fluctuations that most likely reflect measurement noise. On a larger time scale of more than 12 hours, we observed drifts of the trajectories possibly caused by photobleaching or environmental conditions like the density of the cells. On an intermediate time scale we observed changes in nuc/cyt SMAD2 ratio that we interpreted as bursting events in SMAD signaling (cf. Fig.~\ref{fig:data}B). \changed{As shown in ~\cite{Strasene.2018} and S9 Fig, these events cannot be explained by uncorrelated fluctuations in cytoplasmic SMAD.}\rem{R2A2}
    
To quantify how activity bursts vary between cells and TGF-β doses, we developed a method to automatically detect bursting events on an intermediate time scale between 100 and 300 minutes. Our burst analysis consists of three principal steps (see Methods and Fig.~\ref{fig:data}C for details): (1) Slow drifts on a larger timescale are identified and subtracted from the signal.(2) Fast and uncorrelated fluctuations below the relevant time scale of bursting events are removed from the signal. (3) The position, height and duration of the remaining events are detected and quantified. We validated this burst detection algorithm using in silico-generated datasets, in which we on purpose introduced bursts resembling the experimentally observed ones (see Methods).  

This burst detection workflow was applied to experimental single-cell data recorded at low (2.5 pM) and high (100 pM) doses of TGF-β, or in the absence of stimulation (0 pM). In the stimulation datasets, the first signal detected as a burst corresponds to the timing of the initial peak observed in the population average (black line in Fig.~\ref{fig:data}B). The amplitude of these single-cell peaks increased in a dose-dependent manner much like the population-average response (compare Fig.~\ref{fig:data}D). Subsequent activity bursts, enumerated by the burst indices 2-5, reflect stochastic events. These show a substantially lower amplitude than the first burst, but are still elevated in a dose-dependent manner in most cells treated with 2.5 and 100 pM TGF-β when compared to the 0 pM control. Moreover, the total burst count per single cell within the first 24h after stimulation gradually increased with the ligand dose (Fig.~\ref{fig:data}E). In contrast, the duration of the first and subsequent activity bursts remained essentially constant across varying TGF-β doses (compare Fig.~\ref{fig:data}F)\changed{, which agrees to estimations of the signals self-similarity by means of autocorrelations (see S1 Fig and S2 Fig)}. Taken together, these data suggest that certain burst features change in a dose- and time-dependent manner (burst amplitude, total burst count), whereas others apparently do not depend on the external stimulus (burst duration). 

Stochastic burst-like events may contribute to regulating cellular responses to TGF-β stimulation such as cell migration. To quantify the relationship between bursts and cell migration, a motility score for individual cells was calculated as the summed absolute displacement of cells between consecutive time points, as previously described in~\cite{Strasene.2018} (see Methods). By relating the motility of individual cells to bursting events we confirmed that cells with higher SMAD burst counts also tend to exhibit a higher motility than cells with lower burst counts, see Fig.~\ref{fig:data}G. As cells with a lower burst count are also expected to exhibit a lower total SMAD signal, this analysis was performed for various bins summarizing single cells with a similar total area-under-curve (AUC). Even after this correction for the total SMAD signal, cells with a higher total burst count (>12) persisted to exhibit a higher mobility score than those with fewer bursts (<8). This suggest that bursting events in the nuc/cyt SMAD2 ratio are partially correlated to cellular migration and may be causally related to this cellular response.


	\begin{figure}
		\caption{The SMAD-signaling model: from population averages to stochastic single cells}\label{fig:modeling}
	\includegraphics[width=\linewidth]{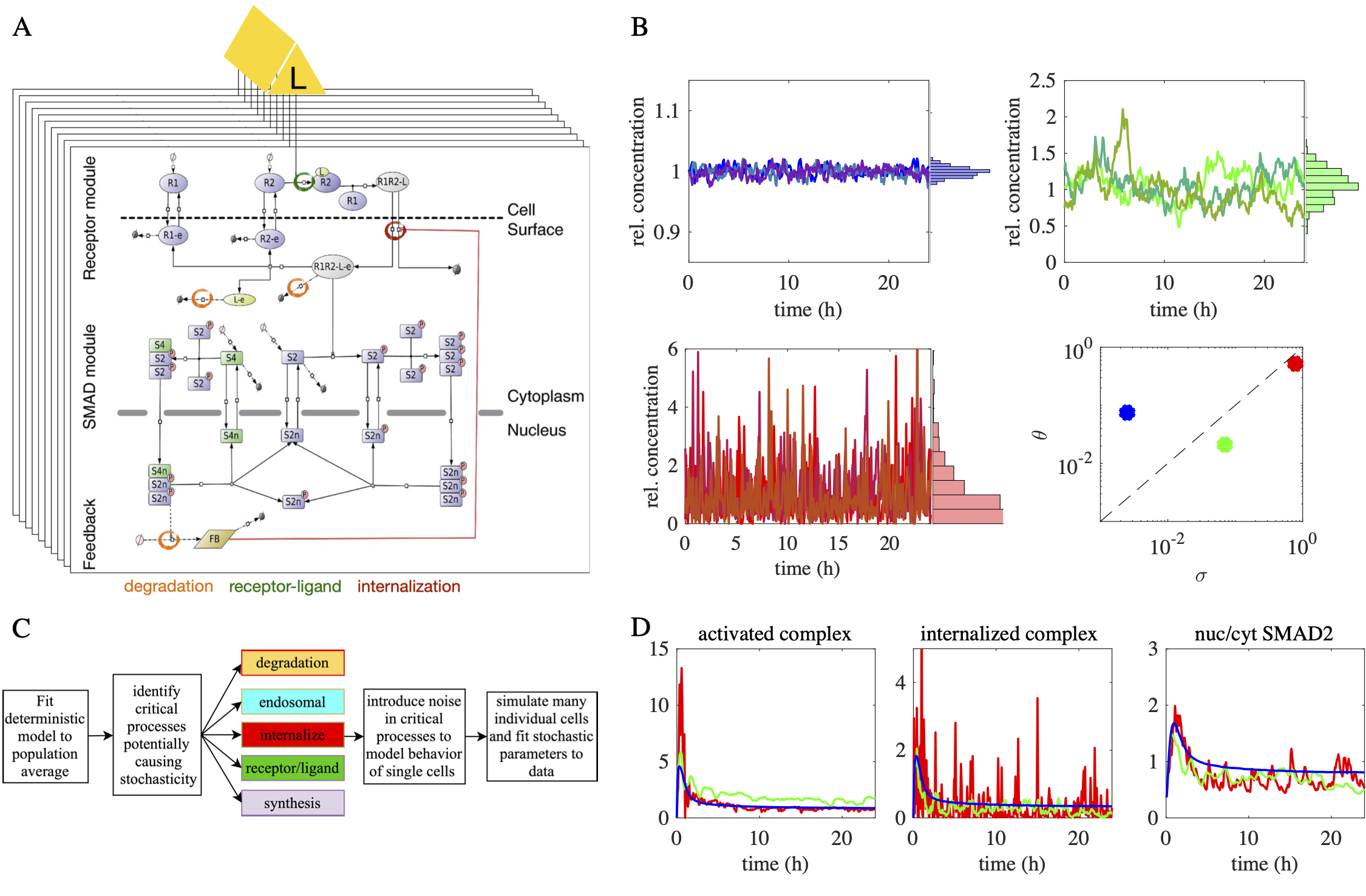}
        \begin{flushleft}A: Topology of the TGF-β pathway model. Extracellular TGF-β (yellow)
		binds to free TGF-β receptors on the cell membrane (blue ovals) to form a receptor-ligand complex (gray ovals). 
		This complex is then internalized into the endosome (R1R2Le) and there functions as 
		an enzyme that phosphorylates SMAD2 (blue rectangle). Phosphorylated SMAD forms homo- and heterotrimers, which are
		transported into the nucleus. Nuclear SMAD further induces the expression of a generic feedback regulator (light green)
		inhibiting TGF-β receptors.
		Extensions in the scheme indicate endosomal (e), nuclear (n) and phosphorylated (p) species. State transitions and intercompartmental shuttling are indicated with
		arrows, enzyme catalysis with circle headed bars, and feedback inhibition with blunt headed
		bars. Colored circles mark parts of the pathway  
		where stochastic noise was introduced in the `degradation', `receptor-ligand' and `internalization' model variants. This panel was modified from~\cite{Strasene.2018}.
		
		B:\ CIR model dynamics for kinetic parameters. Temporal noise in kinetic parameters of the model is shown. For combinations of the variance and reversion parameters $\sigma$ and $\theta$ in the CIR model~\eqref{eq:cir} chosen according to the graph in the bottom right, multiple trajectories are shown, along with their
		temporal distributions, which resemble log-normal distributions. In the top left, the noise parameter
		$\sigma$ was low, whereas it was high
		otherwise, either with a high (bottom left) or low (top right) reversion parameter $\theta$.
		
		C:\ Flowchart of the modeling approach. Starting from the deterministic model, which was fitted
		to the population mean, processes within the signaling pathway (colored reaction steps in panel A) were chosen that are
		sensitive to temporally stochastic occurrences in the cells. Then, random noise (panel B) was
		introduced in
		the kinetic parameters connected to the respective critical process (block model). This
		enabled the reproduction of and model fitting to single cell behavior.
		
		D:\ Propagation of stochastic noise in the model. Noise was introduced in the  receptor internalization rates as done in the stochastic internalization model, see Table~\ref{tab:parameters}. The noise controlling parameters of the CIR model were chosen as indicated by the same colors in panel B. Simulated trajectories of the dynamic model
		concentrations accounting for the activated complex on the cell
		surface ($y_7$, left), the activated endosomal complex ($y_8$, center) and the nuc/cyt
		SMAD2 ratio (right) are depicted. For details on model species see Table \ref{tab:model}. The trajectories show how the temporal noise propagates through the pathway for different parameters in the CIR model.
	\end{flushleft}
\end{figure}

\subsection*{Modeling burst like behaviour in single cells by generalizing a population-average model}

Given the dose-dependent nature of SMAD activity bursts and their potential relation to cell migration, we sought to describe them using a stochastic modeling approach. To this end, we built on a mechanistic model of the pathway published in~\cite{Strasene.2018} that is based on ordinary differential equations (ODEs). This model comprises three modules describing TGF-β receptor dynamics~\cite{Guglielmo2003}, SMAD phosphorylation/complex formation~\cite{Vizan2013} and SMAD-induced feedback regulation~\cite{Legewie2008}, see Fig.~\ref{fig:modeling}A for details.

For a description of bursting, we converted the population-average version of this model into a stochastic description of a complete cell population. This was done by simulating an ensemble of cells, in which each cell exhibits independent stochastic fluctuations in certain kinetic parameter values. The individual cells of this ensemble share a common extracellular TGF-β pool that serves as an input into the single-cell model, but is also in turn influenced by the single cells, as these internalize receptor-ligand complexes and thereby degrade the ligand. Hence, the model consists of interdependent equations describing the extracellular TGF-β and intracellular signaling protein dynamics.

The temporal dynamic of the external input, which is shared among many single cells, is given by
	\begin{equation}\label{eq:lig}
        \changed{dL} = f_L(t, L, \mathbf{Y}^1, \dots, \mathbf{Y}^N, \mathbf{P}^1, \dots, \mathbf{P}^N) \changed{\,dt},
	\end{equation}
	where $L$ denotes the concentration of the free ligand TGF-β and $\mathbf{Y}^i$ is a vector including the dynamic
	concentrations  $y^i_1, \dots y^i_{22}$ relevant to the SMAD pathway in cell $i$ out of $N$ cells. Similarly, the vector $\mathbf{P}^i$ contains the relevant kinetic parameters $p^1_i, \dots, p_{55}^i$ in cell $i$, see Table~\ref{tab:parameters} for details. The right hand side $f_L$ of equation~\eqref{eq:lig} describes the addition of TGF-β into the cell culture dish and its degradation due to internalization into cells. The detailed equations governing the function  $f_L$ are given in Table~\ref{tab:model} \changed{and the dependence on the population size is shown in S12 Fig.}.

The intracellular SMAD signaling dynamics in the i-th cell in our ensemble model are governed by the equation
	\begin{equation}\label{eq:dyn_con}
        \changed{dy^i_k} = f_k(t, L, \mathbf{Y}^i, \mathbf{P}^i) \changed{\,dt}
	\end{equation}
	for $k=1,\dots,22$ species in $i=1,\dots,N$ cells. The right hand sides $f_k$ for $k=1,\dots,22$ account for the reactions inside the cell that depend on the ligand $L$ from the outside, the intracellular dynamical concentrations $\mathbf{Y}^i$ and the kinetic parameters $\mathbf{P}^i$. Refer to Table~\ref{tab:model} for mathematical details.
	
	To account for bursting at the single-cell level, we introduced noise in a subset of kinetic parameters by employing SDEs. This approach leads to different temporal trajectories of the protein concentrations in each cell considered in~\eqref{eq:lig}-\eqref{eq:dyn_con}. We use the Cox-Ingersoll-Ross (CIR) model~\cite{cox1985},
	\begin{equation}\label{eq:cir}
		dp_j(t) = \theta_j(p_j^0-p_j)dt + \sigma_j \sqrt{p_j}d B_j,
	\end{equation}
	to describe the parameter change $dp_j$ at time point $t$ by a reversion force towards the initial value $p_j^0$ and a stochastic contribution due to a Brownian motion $B_j$. The \changed{standard deviation}\rem{R2MA6} parameter $\sigma_j$ controls the time scale of the fluctuations for the corresponding model parameters $p_j$ (see Fig~\ref{fig:modeling}B), while the parameter $\theta_j$ determines the strength of an additional mean reversion force pulling the concentration back to its initial value $p_j^0$. 
	
  	For a suitable model of noise in kinetic parameter we expect first that parameters never fall below zero and thus become nonphysical. Second, we expect temporal stability of the kinetic parameter in the sense that its expectation remains constant and its variance bounded over time. And third, the model should lead to log-normal-like distribution of kinetic parameters over time (compare Figure~\ref{fig:modeling}B) as fluctuations in many cellular processes, e.g., in gene expression, typically lead to this type of distribution~\cite{ozbudakRegulationNoiseExpression2002}.
	
	\delete{Opposed to standard Brownian noise (We compare here to a Brownian noise or Wiener process of the form $dp(t)=\sigma dB$.), the non-negativity of model~\eqref{eq:cir} is verified.}
	\changed{The non-negativity of model~\eqref{eq:cir} is verified (see also Methods for non-negativity of its discretization), although this does not apply to fundamental Brownian motion and the more commonly used Ornstein-Uhlenbeck (OU) model.}\rem{R2MA7}
	Moreover, the model preserves the mean and its variance is bound over time which is unlike Brownian noise and random ODE models whose variance steadily increases with time $t$. We could further verify by simulations that the relative frequency of kinetic parameters over time resembles log-normal distributions (compare Figure~\ref{fig:modeling}B). \changed{Refer also to S3 Fig and S4 Table for a comparison of the fitted CIR model to OU.}\rem{R1A3}
	
    The mean reversion property of the model controlled by the parameter $\theta_j$ leads to burst-like dynamics already in the paths of the kinetic parameters (see Figure~\ref{fig:modeling}B) and our model allowed us to assess how these dynamics propagate to the experimental readout, the nuc/cyt SMAD2 ratio. This ratio is computed from the dynamic concentrations as the sum of nuclear divided by the cytoplasmic SMAD2 species as described in the Methods. The occurrence and strength of fluctuations in the nuc/cyt SMAD2 ratio predicted by the model depend not only on the mean reversion force and variance of the stochastic kinetic parameter. They are also heavily influenced by the  choice of the parameter $p_j$ in which stochasticity is introduced. This sensitivity allowed us to discriminate different model assumptions and identify key parameters for which added noise can reproduce the burst features of the single cell data shown in Fig.~\ref{fig:data}.
    
	\begin{figure}
	\caption{Design of the objective function for the parameter estimation process}\label{fig:fit}
		\includegraphics[width=\linewidth]{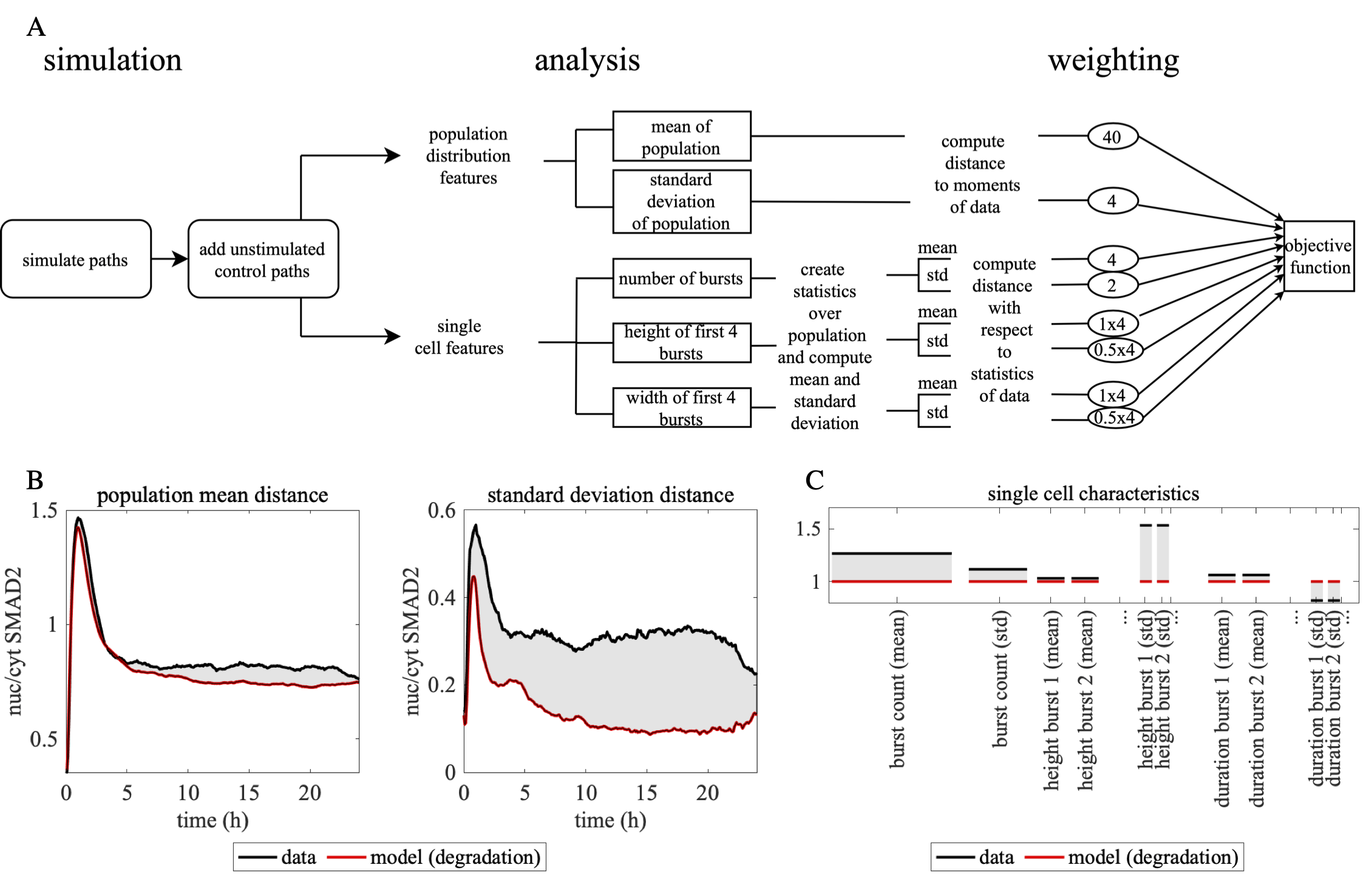}
		
		\begin{flushleft}A:\ Flowchart describing the computation of the objective function for
			parameter estimation. Starting from simulating the nuc/cyt
			SMAD2 ratio in many single cells, to which technical noise in the absence of stimulation was added, the deviation from an experimental data set is computed: The characteristics of
			both population summary statistics  (mean and standard
			deviation of nuc/cyt SMAD2 snapshots over all time points) and individual cells (number of bursts, height and
			width of the first bursts) are included in the objective function. The differences between data and
			simulation with respect to the
			individual components are scaled for comparability and added up with their corresponding weight to obtain a scalar
			value of the objective function. See Methods for details.
			
			B:\ Calculation of snapshot mean and standard deviation as components of the objective function. The population mean (left) and
			standard deviation (right) over time of the nuc/cyt SMAD2
			ratio is shown for an experimental
			data set and a simulation using the degradation model. The gray area between the trajectories reflects
			the corresponding objective function component.
			
			C:\ Bar chart of selected burst related objective function components. The bar chart shows the statistics (mean and standard deviation) for the
			distributions of the single cell properties which are taken into account in the objective function: number of
			bursts, height and duration of the first four bursts. The statistics of the
			data are shown in black and those of the simulation in red. The
			gray area in between corresponds to the weighted component of the statistics as
			taken into account in the objective function. The width of the
			bars reflects the weighting factor.
		\end{flushleft}
	\end{figure}
	
\subsection*{Quantitative model calibration by fitting to single-cell datasets}\label{sec:fit}

By quantitative fitting to experimental data we aimed to reproduce TGF-β-induced bursting in single cells (Fig.~\ref{fig:data}).
To reduce the complexity and to gain insights into the origin of stochastic fluctuations, we considered block models in which we allowed randomness only in sets of parameters describing similar reactions. Technically, each block model uses~\eqref{eq:cir} only for a subset of kinetic parameters $i\in S \subset \{1,2,\dots,55\}$. The remaining kinetic parameters are assumed to be temporally constant. For introducing stochastic fluctuations, we focused on reactions on the receptor level, as receptor dynamics are rate-limiting in TGF-β signaling~\cite{inmanNucleocytoplasmicShuttlingSmads2002, Strasene.2018}. Moreover, we observed in our published single-cell data that complete inhibition of the TGF-β receptors by a small-molecule inhibitor during a stimulation experiment leads to homogeneous decay of SMAD nuclear translocation that showed no sign of stochastic SMAD dynamics at the single-cell level~\cite{Strasene.2018}.

We considered five stochastic block models (see Table~\ref{tab:parameters} for details). The \emph{receptor/ligand} block model introduces noise in the reactions related to the binding of TGF-β to its receptors of type 1 and 2 on the cell surface. The \emph{internalization} block model allows for stochasticity in the transport of receptors and TGF-β/receptor complexes from the cell membrane into the cell. In the \emph{endosomal traffic} block model noise is considered in the intracellular transport of internalized ligand/receptor complexes. Stochastic synthesis of feedback proteins and type 1 and 2 TGF-β receptors is assumed in the \emph{synthesis} block model. Finally, the \emph{degradation} block model adds systematic noise to the degradation rates of the TGF-β receptors and the feedback regulator.

\begin{figure}
  \caption{Validation of path computation scheme and cost function}\label{fig:methods}
    \includegraphics[width=\linewidth]{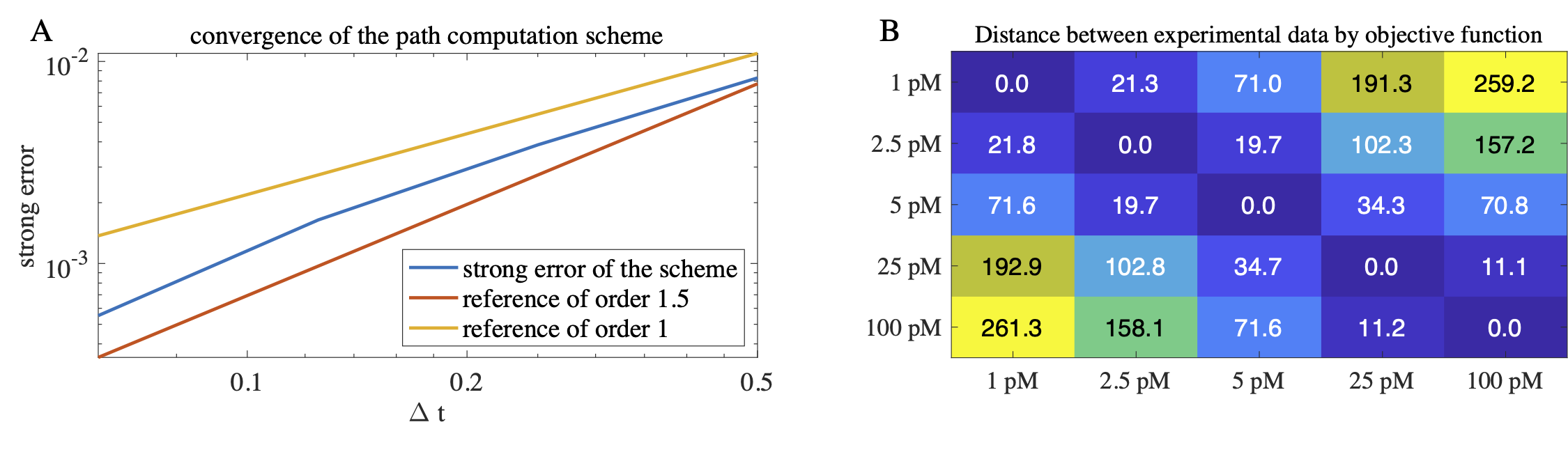}
    \begin{flushleft}
      A:\ Convergence of the path computation scheme. The strong error $|E(X_{\Delta t} - X_{\text{ref}})|$ with respect to the time increment  is shown in loglog scale (blue line). Here $X_{\Delta t}$ refers to the approximate model solution computed by the scheme with time increment $\Delta t$. The reference solution $X_{\text{ref}}$ was computed with a small $\Delta t = 0.03$ and the maximum norm over time was chosen in the error computation. The slope in the graph corresponds to the strong order of convergence. While for larger time increments the scheme exhibits an experimental convergence order only slightly larger than one (yellow line), this order increases to approximately 1.5 for small time increments (red line). Thus, the high order of the scheme is verified (opposed to the commonly used Euler-Maruyama scheme of strong order 0.5) and accurate numerical integration is ensured for small time increments.

      B:\ Sensitivity of the objective function with respect to the dose of TGF-β. Single-cell data from experiments with different doses of TGF-β was compared using our objective function (see Fig.~\ref{fig:fit}). The computed function values (indicated by numbers and color code) show that the objective function increases gradually as the difference in dose increases.
    \end{flushleft}
\end{figure}

To compare the stochastic models to the experiments, we applied the burst detection and analysis used to characterize the data in Fig.~\ref{fig:data} to the model output. This allowed us to define an objective function, which computes a distance measure between simulated and measured cell populations. This function compares summary statistics of the population snapshot distributions (mean and standard deviation of nuc/cyt SMAD2 ratio, Fig.~\ref{fig:fit}B) as well as single-cell characteristics (count, height and duration of bursts, Fig.~\ref{fig:fit}C). Snapshot summary statistics were computed based on the nuc/cyt SMAD2 ratio of all cells at a given time point using the same (5 min) sampling intervals as in the experimental data. As depicted in Fig.~\ref{fig:fit}A, the objective function is a weighted sum of the distance values resulting from the comparison of these single-cell and population statistics (see Methods for further details). The weights were balanced to consider all single-cell features and the standard deviation of the snapshot distribution equally while strictly penalizing deviations from the the snapshot mean through a larger corresponding weight to ensure that the model reproduces the population-average response to TGF-β  stimulation. We validated the objective function by applying it to computationally generated data sets that clearly contained differences in the features we aimed to quantify (not shown). Furthermore, for series of data sets with gradually changing features from a reference data set, we verified that the objective function varied only gradually as well, see Fig.~\ref{fig:methods}B.
	
To calibrate the stochastic models we simulated the nuc/cyt SMAD2 ratio in 375 cells and compared them to the the nuc/cyt SMAD2 ratio measured in 730 cells upon stimulation with 100 pM TGF-β using the objective function. For the simulation of the cells, we implemented a high-order stiff SDE solver that takes into account both the coupling of the ligand degradation by many cells and the stiffness of the system and confirmed its efficiency and accuracy (see Methods and Fig.~\ref{fig:methods}). Interestingly, even in the absence of TGF-β stimulation stochastic events were observed in SMAD nuclear translocation in the experimental data, possibly due to noisy basal shuttling or due to measurement noise. To account for this in the stochastic models, we have added measurement data (normalized by subtracting the mean) from a control experiment without stimulation to the simulated time paths after their numerical computation. As a result even the model version with constant parameters, to which we refer to as `deterministic model', includes temporal noise. 

The kinetic parameters of the deterministic ODE model had been previously fitted to time-resolved measurement data obtained in experiments with different doses of TGF-β in~\cite{Strasene.2018}. With the exception of the kinetic parameters in which we introduced noise, we maintained these originally obtained parameters in our stochastic model. Hence, only the stochastic parts of the models had to be estimated using the objective function. Independent noise was assumed for each stochastic parameter $ p_i $ in a block model, i.e., three values were to be determined per $ p_i $ (the initial value $ p_0^{(i)} $, the noise parameter $ \sigma_i $ and the reversion parameter $ \theta_i $). Depending on the number of stochastic parameters included in a block model between 3 and 9 parameters were estimated during parameter optimization. The extended scatter search~\cite{egeaEvolutionaryMethodComplexprocess2010} within the software package MEIGO~\cite{egeaMEIGOOpensourceSoftware2014} was used to carry out a global parameter search within the parameter space. This evolutionary algorithm uses latin-hypercube sampling to identify an initial population of parameter states, which then evolves by forming combinations between its members. These combination states can replace the original members for the following iteration. The search was stopped when the distance between subsequent populations was lower than a defined level for an extended sequence of iterations. This algorithm together with our high-order path computation scheme could optimize the noise parameters in feasible time.

	\begin{figure}
	\caption{Stochastic internalization model achieves the best fit to the data}\label{fig:results}
		\includegraphics[width=\linewidth]{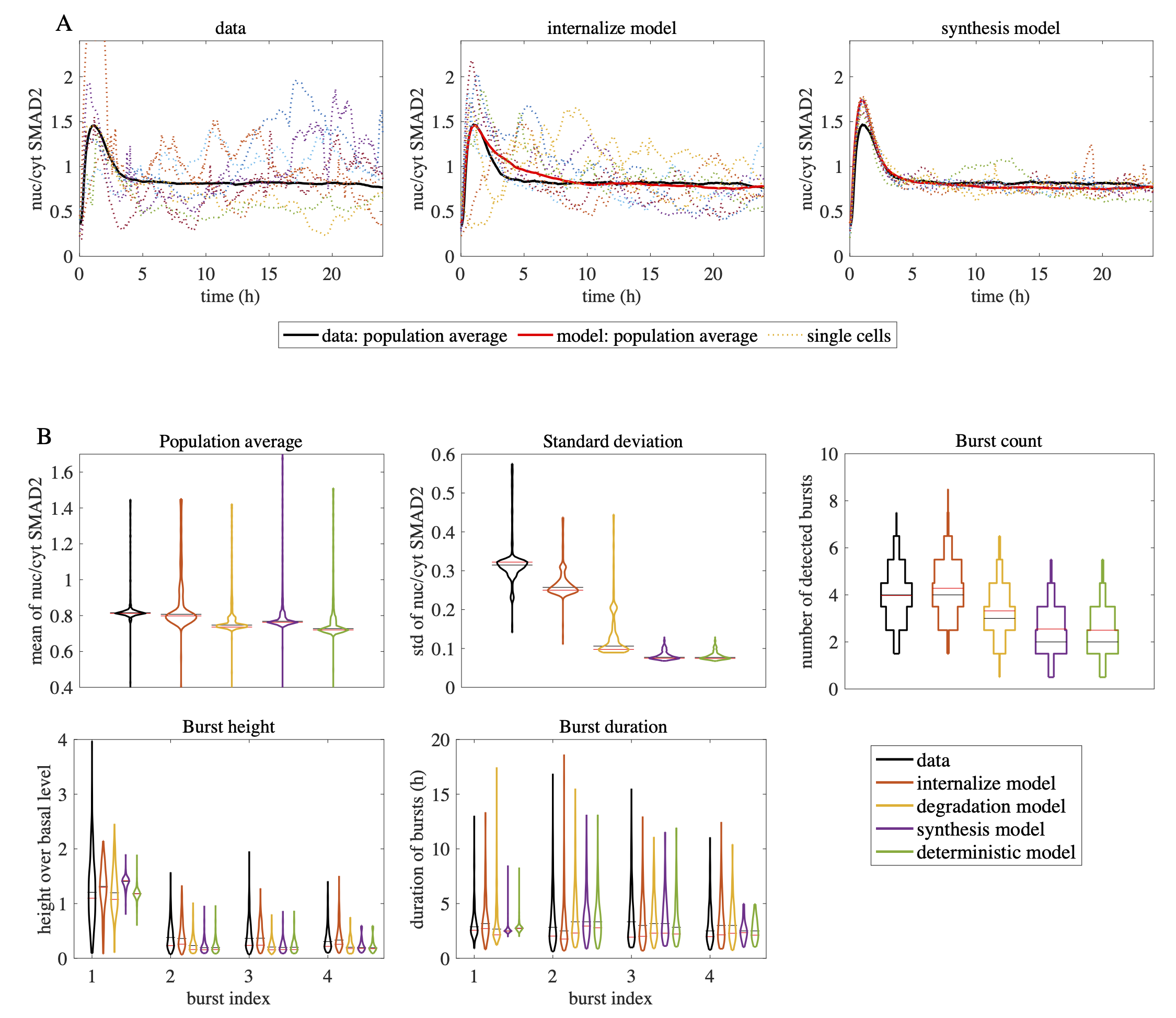}
		
		\begin{flushleft}A:\ Nuc/cyt SMAD2 ratio trajectories of data and stochastic models.
			Experimental data (left), the best-fit internalization model (center) and the best-fit synthesis model (right) are compared for stimulation with 100 pM TGF-β. Examples of single cell trajectories of the nuc/cyt SMAD2 ratio (dashed lines) are shown in addition to the population average over all cells (thick black line). For comparison, the population average of the experimental data (thick red line) is shown alongside with the simulations.
			
			B:\ Population summary and burst feature statistics in the data and stochastic models. Population summary and single cell burst statistics are compared in terms of violin plots for
			experimental data and the stochastic models in case of stimulation with 100 pM TGF-β. The `deterministic' model which introduces noise only through experimental control data (0 pM  TGF-β, see Methods) is additionally considered. The mean and the median are indicated by black and red horizontal lines. Distributions of burst height and duration are compared for the first four detected bursts (burst index). The shown model distributions are based on 5000 simulated cells using the best fit noise parameters. In summary, the internalization model achieved the best fit to the data. \changed{For a comparison with other stochastic methods see S7 Fig and S8 Table.}
		\end{flushleft}
	\end{figure}

\subsection*{TGF-β internalization explains temporal stochasticity in nuclear SMAD}\label{sec:fit}

The optimization results are summarized in Fig.~\ref{fig:results} and Table~\ref{tab:costs}. Distributions of the burst height, width and frequency in all block models are presented in Fig~\ref{fig:results}B.
In Table~\ref{tab:costs} we show the fitting errors of the block models with respect to the objective function. 
In Fig~\ref{fig:results}A various paths of the internalization and the synthesis block models can be compared to the nuc/cyt SMAD2 ratio observed in the experiments.
	
\begin{table}[t]
    \centering
    \caption{Goodness-of-fit of stochastic block models with respect to components of the objective function}\label{tab:costs}
    \begin{tabular}{l@{\hskip .4cm} l l l l@{\hskip .4cm} l l l l @{\hskip .4cm} l @{\hskip .4cm}l l @{\hskip .4cm} r}
      \toprule
      &\multicolumn{12}{l}{model error} \\ \cmidrule(l){2-13}
            model & \multicolumn{4}{l}{burst height} & \multicolumn{4}{l}{burst duration} & count &  mean &  std. & norm\\
      \midrule
      deterministic & 0.12 &     0.20 &    0.50  &  0.57  &  0.10 &     0.03 &    0.33 &   0.44 &    0.87&    2.75 &    5.73 &   11.64 \\ 
      degradation & 0.09&    0.05&   0.14&    0.30 &    0.00 &    0.03 &    0.09&    0.30&   0.38&    0.41&    0.43& 2.21\\
      end.\ traffic & 0.10 &     0.09&   0.17 &    0.28&    0.03&    0.02 &    0.04 &    0.16&    0.17 &    4.62&    3.76 &  9.44 \\ 
      internalization &0.01&   0.00&    0.00 &     0.00 &     0.00 &    0.00 &    0.00 &    0.01 &    0.02 &    1.19 &    0.14 &   1.39 \\ 
      receptor/ligand & 0.09&    0.10&    0.23&  0.33&    0.04 &    0.02&    0.06&   0.22 &    0.26 &    3.66&    3.79 &  8.80 \\
      synthesis & 0.12&    0.19&    0.47 &    0.54 &    0.12 &    0.03&   0.26 &    0.40&    0.76&    1.33&     5.67& 9.90 \\ \bottomrule\\
    \end{tabular}
			
    \begin{flushleft} Distance measurements achieved by the five block models after fitting to
      experimental data for a stimulation with 100 pM TGF-β.
      The objective function with final value shown in the last column is the weighted sum of mean and standard deviation of the nuc/cyt SMAD2 ratio, as well as burst characteristics in single cells including count, height and duration, for details see Fig.~\ref{fig:fit} and Methods.
	  Distance measures for burst height and duration are shown for the first four bursts in each model.
	  In case of single cell statistics the table shows summed contributions of median and standard deviation.
	  For the fitted models no computational instabilities occurred and hence the component accounting for failed simulations is zero.
	  The stochastic internalization model achieved the best fit.
	  \changed{An alternative validation in terms of the Akaike information criterion led to consistent conclusions, see S5 Table.}\rem{R1A4}
      \end{flushleft}
  \end{table}

In summary, our results show that the internalization block model achieved the best fit to the data (compare Fig.~\ref{fig:results}A). All tested models succeeded in being consistent with the population average of the nuc/cyt SMAD2 ratio. However, in terms of temporal stochastic fluctuations significant differences between the models were observed: The block models endosomal traffic, receptor-ligand and synthesis could barely achieve any variability in the observable unlike the internalization and degradation block models (Fig.~\ref{fig:results}B). Moreover, the latter two models attain single cell statistics similar to those of the data, while the other models underestimate the burst count and the variance in their height and duration. 

As reductions of the full complexity of intracellular processes, the stochastic models cannot explain the full variability in the data. Yet, the results are accurate enough to discriminate different models. Our model comparison implies that stochastic processes in the receptor-ligand internalization may play a key role in the temporal heterogeneity of the nuc/cyt SMAD2 ratio in the cell. We hypothesize that the predicted stochastic internalization may arise from low receptor numbers on the cell surface combined simultaneous sorting of dozens of receptors into internalization vesicles~\changed{\cite{Yakymovych2018,Wakefield1987}}\rem{R2A4}. Given that TGF-β receptor become only activated after internalization, the formation of a vesicle will give rise to a sudden increase in the signal and thus to a burst in the nuc/cyt SMAD2 ratio (see Discussion).
		
\begin{figure}
  \caption{Model accurately predicts stochastic behavior at low TGF-β concentrations}\label{fig:prediction}
    \includegraphics[width=\linewidth]{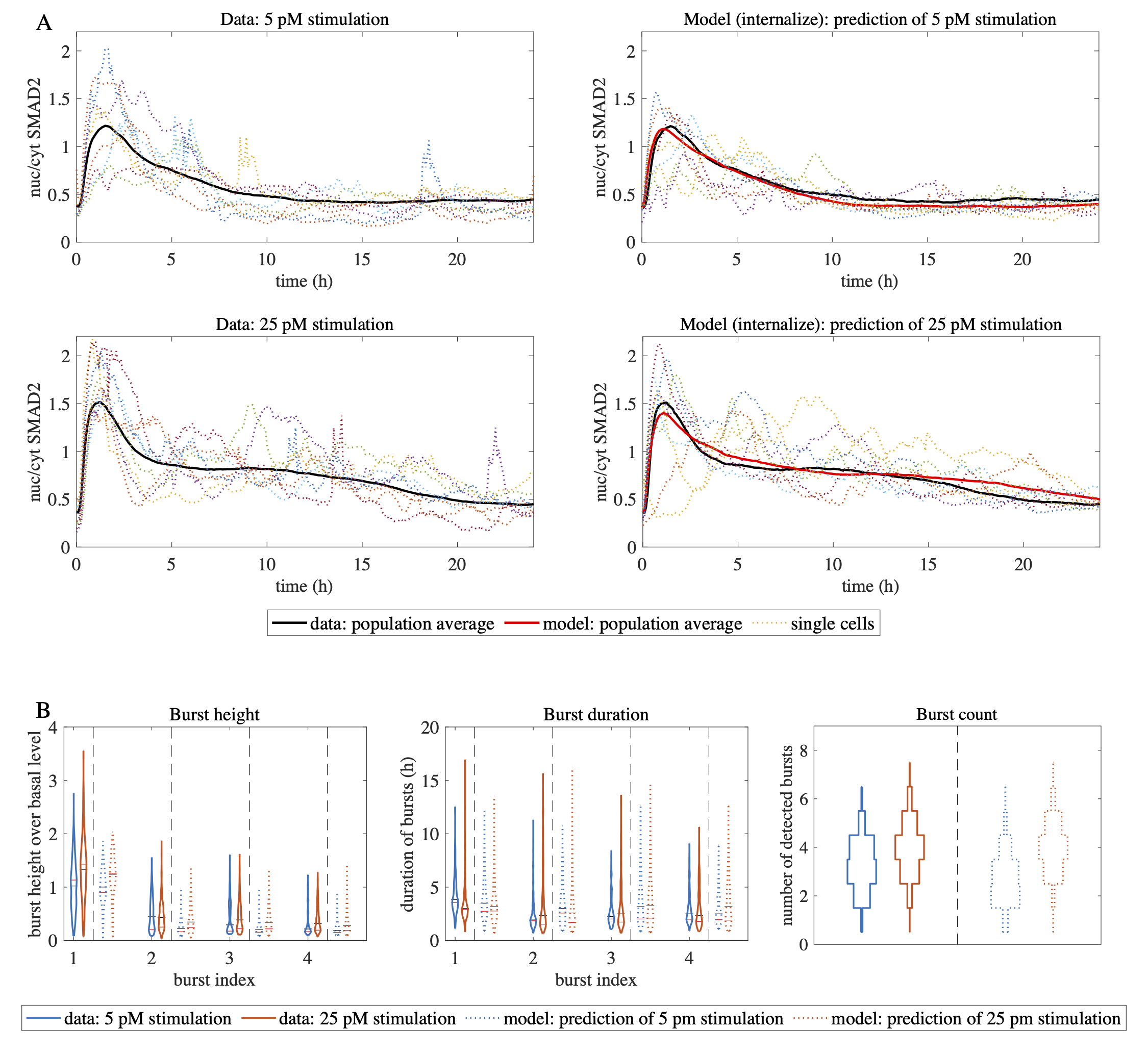}
    \begin{flushleft}A:\ Nuc/cyt SMAD2 ratio trajectories of data at low TGF-β doses and corresponding forecast by stochastic model. Single cell and population average of nuc/cyt SMAD2 ratio of the 5 pM and the 25 pM stimulation experiment are shown in the data (left) and the prediction by the best-fit internalization model (right).
	
	B:\ Distributions of burst features in data and forecast by stochastic model. Single cell burst
	statistics are compared in terms of violin plots for the experimental data
	(solid lines) and the forecast of the internalization model (dashed lines) in case of
	stimulation
	with 5 and 25 pM TGF-β. The mean and the median are indicated by black and red horizontal lines.
      \end{flushleft}
  \end{figure}
	
Stochastic bursting of SMAD2 is a dose-dependent phenomenon, as the burst number and amplitude are reduced at low TGF-β concentrations, whereas the burst duration essentially remains constant (Fig.~\ref{fig:data}). Since the stochastic model has been calibrated solely on the 100 pM TGF-β stimulus data, we asked whether it could successfully predict changes in bursting dynamics at 5 and 25 pM TGF-β, respectively. Indeed, we found that the internalization model also generated realistic predictions of population average and single cell responses for these doses, which had not been employed for model fitting (see Fig.~\ref{fig:prediction}). 
	
Since the stochastic parameters were not fitted to the new stimulation doses, these predictions show the generality of the model. They are also an indication that the stochasticity within TGF-β internalization is an intrinsic property independent of the strength of the signaling. In other words, the processes that lead to internalization noise do not depend on the stimulation dose and the dose mainly influences how strongly the internalization noise is transferred to the nuc/cyt SMAD2 ratio.

\begin{figure}
  \caption{Contribution of temporally stable and unstable noise in restimulation experiments}\label{fig:restimulation}
    \includegraphics[width=\linewidth]{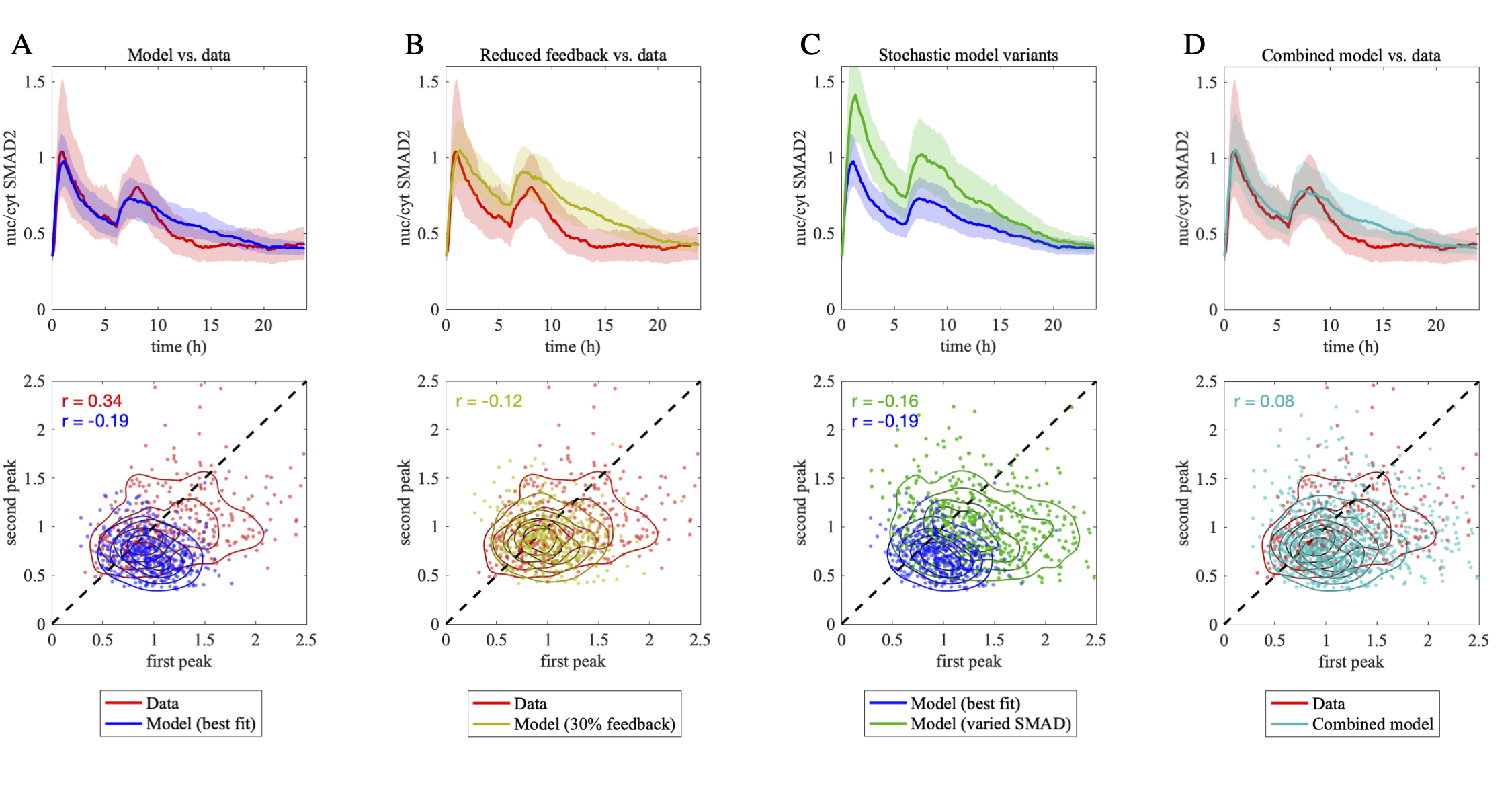}
    \begin{flushleft}

      A:\  Best-fit model fails to fully reproduce the restimulation response. Top: Population median of the nuc/cyt SMAD2 ratio in a restimulation experiment where cells were treated with 5 pM of TGF-β at experiment onset and after 6 hours. Experimental data (red) was compared to the best-fit internalization model (blue). Shades indicate the 25 to 75 \% quantiles. Bottom: Scatter plots comparing the height of the first peak (x-coordinate) with the height of the second peak (y-coordinate) in single cells, each dot representing one cell. The height of each peak in single cells was defined as the nuc/cyt SMAD2 ratio at the peak times in the population-average (55 minutes and 8h, respectively). Contour lines show the two dimensional densities of the cell population. The first and the second peak show a positive correlation in the data (Pearson correlation r=0.34) but a negative correlation in our simulation (Pearson correlation r=-0.19).
      
      B:\ Model with reduced feedback strength shows less pronounced refractoriness to restimulation.  Same as in A, but the experimental data is compared to the model with feedback induction ($P_{39}$ in Table~\ref{tab:parameters}) reduced by 70 \% (yellow). Compared to the orginal model (A), the population-average is increased, especially for the second peak, and the single-cell peak correlation is closer to 0 (Pearson correlation r=-0.12). Both indicate a reduced refractoriness to restimulation. 

      C:\ Minor reduction in the initial SMAD4 level has pronounced effect on stochastic model behavior. The best-fit internalization model is compared to a model variant in which the total SMAD4 concentration is homogeneously reduced by 40\% in all cells of the stochastic ensemble (dark green). A reduction in SMAD4 increases the nuc/cyt SMAD2 ratio by indirect effects on negative feedback regulation.

      D: A model comprising both temporally stable and unstable noise contributions provides a good match to the restimulation data. Artificial cell populations generated by the best-fit model (blue in panel C) and the model variant with reduced SMAD4 (green in panel C) were merged (cyan) and compared to the experimental data (red). The merged population reproduced the variability of the data and the positive correlation of the peak height (Pearson correlation r=0.08).

    \end{flushleft}
\end{figure}

\subsection*{Temporally stable and unstable noise contributions explain variability in response to repeated stimulation}

The stochastic noise in our model leads to signaling fluctuations with a short memory, as the bursts typically decay within a few hours after their appearance (Figs.~\ref{fig:results}A and~\ref{fig:prediction}A). \delete{In our previous work, we characterized the memory of SMAD signaling fluctuations by analyzing sister cells which show a time-dependent desynchronization of signaling behavior after birth from a common ancestor~[7]. In line with temporal stochastic noise, sisters loose similarity in their SMAD signaling behavior within the first few hours after the division event. However, a temporally stable noise component also exists, as sister cell SMAD signaling remains partially similar long after a division event.}
\changed{As shown in~\cite{Strasene.2018}, SMAD signaling also exhibits a temporally stable noise component, since signaling in sister cells remains similar over longer time scales of more than a few hours after cell division.}
\rem{R2A4}
\delete{This raised the question whether our stochastic model can reflect the memory of SMAD signaling fluctuations over extended time scales of several hours.}

Restimulation experiments, in which the same cell population is subjected to a second TGF-β stimulus 6h after the initial treatment, probe the relative contribution of temporally stable and unstable fluctuations, and are thus an interesting test case for our stochastic model: If the response to the first and second stimuli are highly correlated at the single-cell level, the variability is dominated by temporally stable fluctuations \delete{and the pathway exhibits strong memory}. In contrast, temporal fluctuations \delete{reduce the memory of the system and thus} lead to uncorrelated restimulation responses \changed{and} feedback loops in SMAD signaling additionally contribute\changed{.} \delete{, as they introduce history dependence into the pathway. To test whether the stochastic noise assumed here is sufficient to reflect} \changed{To investigate} the long-term memory of the pathway, we compared the best-fit internalization model to restimulation experiments published in~\cite{Strasene.2018} (see Methods for details).  

The model-data comparison was performed at the level of population-average trajectories (Fig.~\ref{fig:restimulation}, top row) and by relating \delete{signaling} \changed{single-cell} peak amplitudes approximately 90 minutes after first and second stimulus \delete{at the single-cell level}\changed{, respectively} (Fig.~\ref{fig:restimulation}, bottom row). The stochastic model approximates the measured population-average trajectories reasonably well \delete{(Fig.~\ref{fig:restimulation}A). In particular, the model} \changed{and} reproduces that the second peak reaches only approximately 80\% of the magnitude of the first peak, thereby indicating \delete{a period of} \changed{signaling} refractoriness after the initial stimulus \changed{(Fig.~\ref{fig:restimulation}A)}. At the single-cell level, the stochastic model performs less well, as the simulated noise reproduces only 48\% of the measured variance of the first peak (initial stimulation), and 62\% of the variance of the second peak (restimulation). \delete{Thus, the signaling variability early after TGF-b stimulation (compare the height of the first burst in Fig.~\ref{fig:results}B and \ref{fig:prediction}B) is not reflected by the model.}Furthermore, the first and second peaks upon restimulation are negatively correlated in the model (Pearson correlation of r=-0.19), whereas their correlation is positive in the data (r=0.34) (Fig.~\ref{fig:restimulation}A, bottom). Taken together, the best-fit stochastic model fails to fully reproduce the restimulation behavior of the SMAD pathway \delete{,especially at the single-cell level}.  

In our model, SMAD signaling is controlled by \delete{a} \changed{transcriptional} negative feedback \delete{regulator that is 428
transcriptionally induced by nuclear SMAD complexes and inhibits the pathway at the receptor level} \changed{loops} (Fig.~\ref{fig:modeling}A) \delete{. Since negative feedback} \changed{which may} introduce pathway refractoriness and are known to reduce signaling variability~\cite{paulsenNegativeFeedbackBone2011}. \delete{, we argued that} \changed{To investigate whether} the feedback strength may be overestimated in the model \delete{.Therefore}, we lowered the parameter of SMAD-dependent feedback induction to 30\% of its best-fit value \changed{(see also S6 Fig for a prediction of feedback reduction in case of single stimulation)}. In line with a role in pathway refractoriness, the reduction of feedback increased the second peak at the population-average level  (Fig.~\ref{fig:restimulation}B, top) and diminished the negative correlation of first and second peaks at the single-cell level (Fig.~\ref{fig:restimulation}B, bottom). Furthermore, the feedback-reduced model showed a higher cell-to-cell variability of both signaling peaks as expected (Fig.~\ref{fig:restimulation}B, bottom). However, significant model improvements at the single-cell level could not be achieved without strong deviations between model and data at the population-average level (Fig.~\ref{fig:restimulation}B, top). Therefore, feedback strength alone does not account for differences between model and data, and this model variant was not considered further. \delete{This suggests that an overestimation of the feedback strength alone does not account for differences between model simulations and restimulation data.}

In~\cite{Strasene.2018}, we \delete{argued}~ \changed{showed} that \changed{temporally stable} SMAD signaling fluctuations are \delete{partially} caused by \delete{temporally stable} differences of signaling protein expression levels across cells.\delete{Using a deterministic model, in which we randomly sampled the initial protein concentrations to create an ensemble of single cells, we could reproduce the heterogeneity in various SMAD trajectory features including the initial peak amplitude~[7]. However, this modeling approach does not account for temporal bursting.} Therefore, we argued that a combination of stochastic noise and stable signaling protein fluctuations may be required to reproduce the restimulation experiments. \rem{R2A7} 

To assess the effect of varying initial protein concentrations in the best-fit stochastic model, we focused on total SMAD4 levels, as this parameter showed the strongest difference between cellular subpopulations \delete{with distinct SMAD signaling dynamics} in our previous \delete{deterministic modeling approach} \changed{work} \cite{Strasene.2018}. In Fig.~\ref{fig:restimulation}C, we homogeneously decreased the total SMAD4 concentration by 40\% in all cells of the stochastic model ensemble (green). Compared to the best-fit stochastic model (blue), the SMAD4-perturbed model shows a higher population-average response and a higher peak variance across cells (Fig.~\ref{fig:restimulation}C). Hence, minor variations in SMAD4 levels\delete{, as they are expected to occur across single cells of a population,} cause a noticeable change in the behavior of the stochastic model. 

\delete{In order} To combine temporally stable SMAD4 fluctuations with stochastic noise, we constructed a cellular ensemble (Fig.~\ref{fig:restimulation}D, cyan) that consists of two stochastic cell populations, one with the best-fit parameters, and one with a homogeneous 40\% reduction in total SMAD4 levels in all cells (Fig.~\ref{fig:restimulation}C, blue and green). This combined model reproduced the major aspects of the restimulation data including the population-average and the variability of the first and second signaling peaks (Fig.~\ref{fig:restimulation}D). Furthermore, the single-cell correlation between the signaling peaks was slightly positive (r=0.08), and thus more consistent with the experimental data than the previous stochastic model variants (Figs.~\ref{fig:restimulation}A-C). \delete{Interestingly, the two underlying cell populations exhibited a negative correlation (Fig.~\ref{fig:restimulation}C), indicating that the positive peak correlation in Fig.~\ref{fig:restimulation}D mainly arises from the SMAD4-induced shift in the population-average.}

Even though the more realistic simulations proposed in this paper were required to capture all aspects of the restimulation data, these results further support our earlier observation that SMAD signaling involves a temporally stable noise component, likely due to variations in signaling protein levels. In addition, stochastic noise and negative feedback also contribute to the restimulation response, e.g., by dampening the positive correlation of the first and second signaling peaks. Therefore, the pathway responds in a complex and non-deterministic fashion to repeated stimulation. 

In summary, our restimulation analysis helped to assess the interplay of noise sources and feedback mechanisms in determining the long-term heterogeneity of SMAD signaling in single cells. 

\section*{Discussion}

In this work, we employed stochastic differential equations to model temporal variability observed in single-cell SMAD signaling as stochastic fluctuations in kinetic parameter values. Assuming a CIR process in TGF-β receptor internalization, the proposed model realistically reproduced noisy signaling at the single-cell level. To quantitatively calibrate the model based on experimental data, we developed a method to detect and quantify bursting events. In combination with an adjusted high-order time integration scheme, we thereby developed an efficient optimization scheme to discriminate different model variants based on single-cell measurements.
 
In previous modeling studies, the SMAD pathway was mostly described at the population-average level, i.e., the simulations described the dynamics of one average cell \cite{Melke2006,Vilar2006,Schmierer2008,Chung2009,Clarke2009,Celliere2011,Wegner2012,Zi2012,Nicklas2013,Vizan2013,Khatibi2017}. Now that more and more single-cell datasets of SMAD dynamics are available \cite{Zi2012,Frick2017,Li2018,Tidin2019}, quantitative modeling of cell-to-cell variability in the pathway becomes feasible. In~\cite{Strasene.2018}, a first attempt was made by sampling the initial protein concentrations in a deterministic model. Thereby, a heterogeneous ensemble of single cells was simulated and the heterogeneity was assumed to be temporally stable. Here, we have now focused on understanding the source of temporal fluctuations observed in live-cell imaging of SMAD2 nucleocytoplasmic shuttling.

Specifically, we analyzed SMAD2 bursting events which may contribute to variability in the cellular response to TGF-β, since cells with more bursts exhibit higher motility in the cell culture dish, clf. Fig.~\ref{fig:data}G. Apparently, stochastic bursts were previously reported for the nuclear translocation of the closely related SMAD4 protein during the development of Xenopus embryos \cite{Warmflash2012}. Our implementation of automated burst detection allowed us to quantify burst height, width and frequency at the level of individual cells. We found that the number and height of bursting events are dose-dependent, and therefore contain information about the extracellular TGF-β concentration. To describe this bursting phenomenon, we have introduced temporal noise in specific kinetic parameter values and analyzed how this noise propagates to the experimentally observed nuc/cyt SMAD2 ratio, clf. Fig.~\ref{fig:modeling}B and~\ref{fig:modeling}D. 

Quantitative stochastic modeling in biology so far is mostly limited to a description of simple reaction networks, e.g., promoter cycles describing the activity of single genes \cite{Paulsson2005,Raj2008,Molina2013,Fritzsch2018,Loos2019}. Our large-scale signaling pathway model (see Fig.~\ref{fig:modeling}) contains 45 kinetic parameters and therefore poses a challenge to stochastic modeling,  especially when calibrated based on a single-cell dataset with high temporal resolution (0-24h in 5 min intervals) and a large number of cells (n=730). To overcome this challenge, we proposed a hybrid deterministic-stochastic approach where temporal fluctuations were modeled by a CIR process in a subset of the kinetic parameters. In contrast to the Wiener process, which is commonly used in stochastic modeling, the CIR process guarantees non-negative dynamic variables. The properties of the CIR process are well suited for biological modeling of noise: the temporal distribution of the noise tends to a log-normal distribution, clf. Fig.~\ref{fig:modeling}B, which was reported for kinetic rates also from experiments, see for example~\cite{ozbudakRegulationNoiseExpression2002}.

In this work we have implemented a time integration scheme that is capable to achieve reliable simulations and fast enough to fit the complex model to detailed data. The scheme’s efficiency has been obtained by high order and semi-implicit time stepping. The non-negativity of the CIR process has further contributed to the stability during parameter estimation. The identification of appropriate noise parameters for our model required a calibration that also takes properties of the fluctuations into account. Instead of comparing full distributions, which could be achieved for example using the Wasserstein distance, we used the first two statistical moments of the burst feature in our objective function. While the duration of the bursts did not exhibit significant differences among several tested models the burst height and their count played a crucial role in their discrimination, clf. Fig.~\ref{fig:fit}B.

Using our burst detection technique, we could discriminate different models and quantify the contribution of one or several parameters to the observed noise. We showed that the receptor internalization model performed better than the degradation and the synthesis models, clf. Fig.~\ref{fig:fit}. In the synthesis model, we observed low variability close to the results of the deterministic model. The degradation model variability is closer to the experimental data in terms of burst statistics, but is still outperformed by the internalization model. These results indicate that bursts might be caused by stochastic events in receptor internalization. In recent papers studying the internalization of Epo receptors and of the yeast methionine transporter Mup1 by live-cell imaging it has indeed been shown that receptor transport processes can exhibit strong heterogeneity \cite{Kallenberger2017,Stelling2019}. A plausible explanation for noise arising in receptor internalization is that large numbers of receptors on the cell surface get constricted and internalized at once during vesicle formation. Given that TGF-β receptor internalization promotes downstream signaling \cite{Yakymovych2018}, the formation of a vesicle will give rise to a sudden increase in the signal and thus to a burst in the nuc/cyt SMAD2 ratio \changed{(see also S10 Fig. for an alternative internalization model)}. For EGFR receptors, it was indeed observed that each internalized vesicle in the endosome contains around 100 receptors ~\cite{villasenor2015regulation}. Given that TGF-β receptors are present on the cell surface in low amounts of few hundreds to thousand molecules \cite{Yakymovych2018,Wakefield1987} and are internalized via coated pits \cite{ehrlich2001}, these internalization bursts may indeed give rise to significant fluctuations in downstream signaling as we predict here.

\changed{Our findings about the role of receptor internalization in the SMAD pathway suggest promising ways to derive simplified models with similar burst dynamics in the future (e.g. through quantized state systems methods). }\rem{R1A6}

The best-fit internalization model was independently validated by analyzing the bursting behavior at low TGF-β concentrations. The successful prediction of bursting at various TGF-β levels proved that the noise parameters found by fitting to the 100 pM are not specific to this condition. Hence, the internalization noise remained constant across conditions, but its propagation through the signaling network changed, so that the net result is a lower signaling output noise in terms of burst count and amplitude upon weak TGF-β stimulation (Figs.~\ref{fig:results} and~\ref{fig:prediction}).  

The restimulation experiments showed that the stochastic model only partially accounted for the pathway behavior upon repeated TGF-β application (Fig.~\ref{fig:restimulation}): In the experimental data, the first and second signaling peaks are positively correlated, whereas the model, which focuses on temporally unstable noise, predicted a negative correlation between peaks. By performing detailed simulations, we could show that correlated restimulation behavior is explained by the stochastic model if we additionally assumed a temporally stable noise contribution. As in our previous work \cite{Strasene.2018}, this temporally stable noise contribution was implemented as cell-to-cell differences in the total signaling protein concentrations which are important determinants of signaling fluctuations in various pathways \cite{Feinerman2008,Spencer2009,Kallenberger2014,Karlsson2015,Kallenberger2017,Dixit2020,Sarma2020}. The different time scales of these two noise contributions explain why sister cells rapidly desynchronize their SMAD signaling behavior after a division event (temporally unstable noise), but remain partially similar over long time scales (stable noise) \cite{Strasene.2018}. Likewise, the single-cell correlation of the first and second peaks in a restimulation experiment will likely become more and more dissimilar if the second treatment is delayed relative to the first one. Our simulations have shown that stochastic modeling is a valuable tool for the analysis of such experiments which provides insights into the underlying mechanisms.  

To the best of our knowledge, this work is the first approach to model bursting events in signaling pathways using CIR processes in kinetic parameters. Our results demonstrated that this approach is suitable to simulate single-cell dynamics and to analyze the origin of noise in large systems. The optimization framework combining burst detection and an efficient time integration scheme are not limited to TGF-β signaling and can be generalized to other systems in the future.

\section*{Methods}
\subsection*{Single-cell data}
The experimental datasets used in this paper are published in~\cite{Strasene.2018} and are available online at the Dryad digital repository~\cite{Strasene.2018DATA} and in the GitHub repository~\cite{stochSMADGit}. In brief, human MCF10A cells expressing SMAD2 fused to the yellow fluorescent protein mVenus and H2B fused to the cyan fluorescent protein mCerulean under the control of the Ubiquitin C promoter were imaged on a Nikon Ti inverted fluorescence microscope with a CCD camera and a 20x plan apo objective using appropriate filter sets. The microscope was surrounded by a custom enclosure to maintain constant temperature and atmosphere. Images were acquired every 5 minutes for the duration of the experiments. Cells were tracked in the corresponding images using custom-written scripts and fluorescent intensities quantified. For parameter estimation of the stochastic models (Fig.~\ref{fig:restimulation}), the measured nuc/cyt SMAD2 ratio after stimulation with 100 pM of TGF-β from the published dose-titration experiment (Fig.~\ref{fig:modeling}A-D in~\cite{Strasene.2018}) was merged with a second, previously unpublished dataset with the same experimental setup (N = 378 cells). For model validation (Fig.~\ref{fig:prediction}), we used the nuc/cyt SMAD2 ratios from the 5 pM and 25 pM conditions from the same dose titration experiments. In the restimulation analysis (Fig.~\ref{fig:restimulation}), the experimental data from repeated 5 pM TGF-β stimulation (Fig. 4F in~\cite{Strasene.2018DATA}) was used.

\subsection*{Burst analysis}

To successfully apply burst detection to experimental data, we tuned the parameters of the above-mentioned processing steps, such as the bandwidth of the Gaussian filters and the height thresholds of the peak detection. To create an in silico benchmark for burst detection, we generated synthetic trajectories by model simulations and added Gaussian hills as artificial burst events as well as fluctuations from unstimulated cells as noise. We chose the burst detection parameters such that in silico added bursts are detected reliably (high sensitivity: high true positive detection rate), while keeping the detection of bursts in unstimulated cells low (high specificity: low false detection rate). Specifically, we randomly sampled all burst detection parameters and calculated sensitivity and specificity for each burst parameter candidate. Thereby, a receiver operating characteristic (ROC) curve was generated~\cite{powersEvaluation2011} and the best parameter set achieved a false detection rate below 25\% and a sensitivity of more than 75\% (data not shown).
	
\subsection*{Objective function}
The objective function that we used for the model validation and parameter estimation is a weighted sum of several distance measures accounting for single-cell characteristics and summary statistics of the population distribution, see also Fig.~\ref{fig:fit}.  All these distance measures were calculated based on the SMAD2 nuc/cyt ratio which in the model is given by
\begin{equation}\label{eq:ratio}
		r = \frac{y_{18} + 2 y_{19} + 3y_{20} + y_{21}}{y_{10}+ y_{11} + 2y_{13} + 3 y_{14}}.
              \end{equation}
Hereby, we refer to Table 2 for details on the dynamic variables employed.

\paragraph{Single cell characteristics}
	To quantify the fit in burst height and duration we considered the relative difference between model and data in median and standard deviation across simulated and measured cell population. In more details, consider the empirical population $X$ including the nuc/cyt SMAD2 ratio in $N_{X}$ cells either given through measurement data or the model. We define the distribution of the height of the $i-$th burst computed with our burst detection by $p_\text{height}^i(X)$. We note that a)~the burst detection might detect less than $i$ bursts and b) in case of a model population the path computation might fail to compute some of the paths. Hence we assume that $p _\text{height}^i(X)$ considers the $i$-th burst height only if a) and b) can be excluded. Moreover we define $r^i(X)\geq0$ to be the relative number of paths where either a) or b) holds for the $i$-the burst deducting the relative number of paths where a) holds in the data. By denoting measurement data by $Y$  and model results by $X$ we set
	\begin{equation}\label{eq:heightMed}
	  d^{i, \text{med}}_{\text{height}}(X,Y) = r^i(X) + \left( 1 - r^i(X) \right) \frac{|\tilde p_{\text{height}}^i(Y) - \tilde p_{\text{height}}^{i}(X)  |}{|\tilde p_{\text{height}}^i(Y) + \tilde p_{\text{height}}^{i}(X) |},
	\end{equation}
	where $\tilde p$ denotes the median of the distribution $p$. Hence, the distance measure increases if the burst height is different between model and data, and/or if a) or b) holds for a large number of cells. Similarly, we quantify the distance in terms of standard deviation
	\begin{equation}\label{eq:heightStd}
	  d^{i, \text{std}}_{\text{height}}(X,Y) =  r^i(X) + \left( 1 -  r^i(X) \right) \frac{|\sigma ( p_{\text{height}}^i(Y)) - \sigma(p_{\text{height}}^{i}(X)) |}{|\sigma( p_{\text{height}}^i(Y)) + \sigma(p_{\text{height}}^{i}(X)) |},
	\end{equation}
	with $\sigma(p)$ referring to the standard deviation of the distribution $p$. Denoting by $p_\text{dur}^i(X)$ the distribution of the duration of the $i-$th burst in an empirical population $X$, we can define the distances $d^{i, \text{med}}_{\text{dur}}(X,Y)$ and $d^{i, \text{std}}_{\text{dur}}(X,Y)$ in the same way and obtain the corresponding formulas by exchanging the distributions in \eqref{eq:heightMed} and \eqref{eq:heightStd}. The distance measures with respect to both burst height and burst duration are considered for the first four measured bursts and hence, including median and standard deviation, they account for 16 components in the objective function.

The distance with respect to the burst count is, similarly as their height and duration, measured in terms of relative difference in mean and standard deviation. Denoting by $p_{\text{count}}(X)$ and $p_{\text{count}}(Y)$ the distribution of detected bursts in the model and in the data we computed distance measures in mean and standard deviation like in~\eqref{eq:heightMed} and~\eqref{eq:heightStd} and included them in the objective function. In the computation we took the relative number of failed path computations, $r^0(X)$, into account. To avoid parameter ranges that cause instabilities during the parameter estimation we included this rate as a separate component increasing the value of the objective function.       

  \paragraph{Population mean and standard deviation}
	The mean and standard deviation of the nuc/cyt SMAD2 ratio across the simulated and measured cell population account for two further components of the objective function. At any fixed point in time these quantities can be computed employing the snapshot distribution across all cells. Thus, for populations $X$ and $Y$ time-dependent means and variances $m_X^t, m_Y^t, \sigma_X^t, \sigma_Y^t$ can be considered. The global distances of the populations with respect to mean and standard deviation sum the differences over time and take relative values using the formulas
	$$d^\text{mean}(X,Y) = \frac{\sum_t |m_X^t-m_Y^t|}{\sum\limits_t  |m_X^t|+\sum\limits_t
	|m_Y^t|}, \quad  d^\text{std}(X,Y) = \frac{\sum_t |\sigma_X^t-\sigma_Y^t|}{\sum\limits_t  |\sigma_X^t|+\sum\limits_t
	|\sigma_Y^t|}.$$
	
      \paragraph{Weighting} A balanced weighting of the introduced distance measures with respect to single-cell characteristics and population statistics allowed us to compare measured and simulated populations in terms of their response to TGF-β stimulation.  In particular, we considered the height and duration of the first four bursts and weighted burst height, burst duration, burst count and standard deviation of the observable equally. To prevent deviations from the population mean and failed path computations in the parameter estimation we assigned large weights to the corresponding components. Moreover, we weighted differences in median and mean higher than differences in standard deviation. 

	In more detail, for measurement data $Y$ and model results $X$, we defined the output of the objective function as the vector
	\begin{align*}
	  d(X,Y)= \left(\right. &d^{1, \text{med}}_{\text{height}}(X,Y), \dots, d^{4, \text{med}}_{\text{height}}(X,Y),  \frac12 \, d^{1, \text{std}}_{\text{height}}(X,Y), \dots, \frac12 \,d^{4, \text{std}}_{\text{height}}(X,Y), \\
	& d^{1, \text{med}}_{\text{dur}}(X,Y), \dots, d^{4, \text{med}}_{\text{dur}}(X,Y),  \frac12 \, d^{1, \text{std}}_{\text{dur}}(X,Y), \dots, \frac12 \,d^{4, \text{std}}_{\text{dur}}(X,Y),\\
   & 4 \, d^{\text{mean}}_{\text{count}}(X,Y), 2 \, d^{\text{std}}_{\text{count}}(X,Y), 80 \, r^i(X), 40\, d^{\text{mean}}(X,Y), 4 \, d^{\text{std}}(X,Y) \left. \right),
	  \end{align*}
	and considered as scalar distance the sum over its squared components.

	\subsection*{Path computation}
	For efficient and accurate numerical approximation of the model~\eqref{eq:lig}, \eqref{eq:dyn_con}, \eqref{eq:cir} we employed a method based on the Itō-Taylor formula of strong order
	$\gamma=1.5$ from~\cite{HanKloeden2017}. The formula includes the control parameter $\theta$, which we chose $\theta=1/2$ resulting in a semi-explicit scheme that can both handle the stiff reaction terms of the SMAD signaling model and resolve the dynamics accurately
	without the need to compute second-order derivatives.
	Its deterministic counterpart is known as the Crank-Nicolson-method and commonly used for parabolic differential equations.
	The scheme in its full form reads
	\begin{align*}\label{eq:scheme}
		L_{n+1} = L_n + &\frac{\Delta t}{2} f_L(t_{n+1}, L_{n+1}, \mathbf{Y}^1_{n+1}, \dots, \mathbf{Y}^N_{n+1}, \mathbf{P}^1_{n+1}, \dots, \mathbf{P}^N_{n+1}) \\ +  &\frac{\Delta t}{2} f_L(t_{n}, L_n, \mathbf{Y^1_{n}}, \dots, \mathbf{Y}^N_{n}, \mathbf{P}^1_{n}, \dots, \mathbf{P}^N_{n})\\
				y^i_{k,n+1} = y^i_{k,n} + &\frac{\Delta t}{2}\left[ f_k(t_{n+1},  L_{n+1}, \mathbf{Y}^k_{n+1}, \mathbf{P}^k_{n+1}) +  f_k(t_{n}, L_{n}, \mathbf{Y}^k_{n}, \mathbf{P}^k_{n})\right] + G^{i}_{k,n}\\
		p^i_{j, n+1} = p^i_{j, n} + &\frac{\theta_k\, \Delta t}{2}\left[ 2 p^i_{j, 0} - p^i_{j, n+1} -p^i_{j, n} \right] + \sigma_j \sqrt{p^i_{j, n}} \Delta W^{i}_{j,n} \\ & + \frac{\sigma_{j}}{4}\left[ (\Delta \changed{B}^{i}_{j,n})^{2} - \Delta t\right ] + H^i_{j,n},
	\end{align*}
	for $i=1,\dots, N$, $k=1,\dots,22$ and $j\in S$. Here $L_{n}$, $y^i_{k,n}$ and $p^i_{j, n}$ denote approximates of the ligand, the dynamic concentrations and the stochastic kinetic parameters at time instance $t^n=n \Delta t$. Moreover, $\Delta t$ refers to the time increment and $\Delta \changed{B}_{j,n}^i = \sqrt{\Delta t} \mathcal{M}_{j,n}^i$, where $\mathcal{M}_{j,n}^i$ are independent and identically distributed normal random variables with expectation zero and variance one. The high-order correction terms $G^{i}_{k,n}$ and $H^{i}_{j,n}$ require the generation of further random variables and its formulas are given in Table~\ref{tab:highorder}.
	
	To solve the nonlinear systems in each time step, we used an adaptive Newton method, where the Newton-error is individually computed for each cell. If this error fell below the error tolerance the corresponding cell was removed from the system in the next Newton iteration. This was iterated until convergence was achieved in each cell. This technique significantly reduced the computational costs. We optimized the performance further by employing a sparse Jacobian that we symbolically pre-computed offline. 
To handle the stiff reactions caused by the step-like addition of the TGF-β stimulus we employed small time increments of size $\Delta t = 0.04$ in the stimulation phase, which included only a minor part of the total time interval. For the remaining time steps we switched to $\Delta t = 0.6$ for fast and robust computations.
	
Although our model leads to non-negative protein concentrations and kinetic parameters, the scheme cannot guarantee non-negativity. To avoid nonphysical parameters in simulations we applied clipping and set computationally obtained negative state variables to zero. Further, in the case of kinetic parameters equal to zero we ignored the high-order corrections in the scheme for the next time step. We experimentally verified the strong convergence of the resulting scheme and observed a strong order higher than 1, see Fig.~\ref{fig:methods}A.
	
\subsection*{Simulations of restimulation experiments}	

In restimulation experiments (Fig.~\ref{fig:restimulation}), the system is subjected twice to a 5 pM TGF-β stimulus at 0 and 6h. To simulate these experiments, the concentration of external ligand in the model was raised from 0 to 5 pM at the beginning of the experiment, and from the value at the time point before 6h to 5 pM at t = 6h. The restimulation experiment was performed using a different charge of recombinant TGF-β when compared to the dose-titration experiments used for model fitting and validation (Figs.~\ref{fig:results} and~\ref{fig:prediction}). Thus, the initial peak amplitudes of the signal 1h after stimulation were slightly different between both types of experiments (compare Figs.~\ref{fig:prediction}A and~\ref{fig:restimulation}A), likely because different ligand charges have slightly different biological activities. To account for this ligand differences in our simulations, the forward and backward reaction rates of ligand binding ($P_{49}$ and $P_{20}$, clf. Table~\ref{tab:parameters}) were adapted and a dose factor changing the effective external ligand concentration was introduced. To avoid an over-adaptation, only data-points the second stimulus in the restimulation experiments were used in the estimation of these ligand-adjustment parameters.

\section*{Supporting information}
\changed{
\paragraph{S1 Fig. Autocorrelation of nuc/cyt SMAD2 ratio}
Autocorrelation functions of the nuc/cyt SMAD2 ratio for both the experimental data and the respective simulations (after subtracting the trend) for all 5 considered doses. Autocorrelations for model and data have similar trajectories and reach values close to 0 at similar times fitting within the range considered as intermediate time scale in our definition of bursts (grey shaded area). \url{https://doi.org/10.6084/m9.figshare.19064561}
 
 \paragraph{S2 Fig. Autocorrelation of dynamic variables in the model}  Autocorrelations of individual species within the pathway in the model after subtracting time averages in case of stimulation with 100 pM TGF-β. The autocorrelation trajectories vary among each other, they mostly drop fast in the first 120 minutes but then approach 0 only slowly. In species with slower decay of self-similarity like nuclear and cytoplasmic unphosphorylated SMAD2, the noise led to faster decay. In species with faster decay of self-similarity on the other hand, like nuclear phosphorylated SMAD2 and heterotromer, it led to slower decay. In both cases the stochasticity brings the signal closer to the autocorrelation of the nuc/cyt SMAD2 ratio. No species alone shows an autocorrelation similar to the one of the SMAD2 nuc/cyt ratio (compare S1 Fig).
 \url{https://doi.org/10.6084/m9.figshare.19064630}

\paragraph{S3 Table. Comparison between the CIR and the OU model}
Distance measurements of the OU internalisation model in comparison to the CIR internalisation and the deterministic model to experimental data for a stimulation with 100 pM TGF-β (compare Table~\ref{tab:costs}). The best-fit parameters from fitting the CIR internalisation model were used in the OU internalisation model. Parameters could not be a priori estimated for the OU internalisation model due to instabilities caused by negative stochastic parameter values. The components of the objective function in the OU internalization model are consistent and comparable albeit of an increased magnitude in comparison to the results of the CIR internalization model.
\url{https://doi.org/10.6084/m9.figshare.19064483}

\paragraph{S4 Fig. Best-fit OU model}
Single cell and population average of nuc/cyt SMAD2 ratio for 100 pM TGF-β stimulation predicted by the OU internalisation model (see S2 Table). In comparison to the CIR internalisation model qualitatively similar bursting is observed in the single cells (compare Fig.~\ref{fig:fit}A).
\url{https://doi.org/10.6084/m9.figshare.19064663}

\paragraph{S5 Table. Log-likelihood and Akaike information criterion of block models}
In analogy to simultaneous fitting of an error model proposed in ~\cite{raue2013lessons}, the scaling of the residuals of our cost-function was re-normalized by introducing the rescaling factors $s_k$ for the components of the objective function $r_k$ minimizing 
$\sum_{k} (r_k/s_k)^2 + 2\, \log(s_k)$. 
Factors for all components referring to burst height and those referring to burst duration were assumed equal, respectively.
This re-scaled cost-function met the requirements on the variance of the residuals to calculate the log-likelihood (LL) and the Akaike information criterion (AIC) of the block models in case of the 100 pM TGF-β stimulation. The AIC ratio was computed with respect to the AIC of the internalization block model and indicates that the internalization model outperforms the other models also taking the degrees of freedom into account. \url{https://doi.org/10.6084/m9.figshare.19064558}

\paragraph{S6 Fig. Prediction of SMAD7 knockout}
   SMAD7 knockout prediction by reducing the parameter of SMAD dependant feedback induction in the internalization model to 30 \% of its original value in case of a single stimulation with 100 pM TGF-β. Using burst analysis, the model predictions were compared to data from a SMAD7 knockout experiment. Simulation data and burst analysis is presented analog to Fig.~\ref{fig:results}. The population average is well described by the model prediction. 
  Furthermore, the model predicts a slight increase in the mean burst count, amplitude and duration in the SMAD7 knockout. Moreover the burst amplitude and duration variance is predicted to increase. Most of these predictions are indeed observed qualitatively in the SMAD7 knockout compared to wildtype data, although the effect sizes are not very large. The prediction was similarly accurate to those of 5 pM and 25 pM doses of TGF-β (compare Fig.~\ref{fig:prediction}). 
 \url{https://doi.org/10.6084/m9.figshare.19064687}

\paragraph{S7 Fig. Comparison to full and hybrid stochastic simulation techniques}
We applied $\tau$-leaping SSA and a hybrid SSA solver to the TGF-β model in case of stimulation with 100 pM TGF-β using the COPASI toolbox \cite{copasi2006} and the model from \cite{Strasene.model.2018}.
In a direct comparison between $\tau$-leaping SSA and the internalization CIR model by running 375 paths on the same hardware, the CIR internalization model performed nearly 100x faster (2.2 sec/path vs. 210 sec/path). The hybrid approach combined SSA with a Runge-Kutta method and selected the stochastically treated species based on the number of particles, which also correlates with the expected variance. The publicly available hybrid solver has been much slower than our approach. Simulation data and burst analysis is presented analog to Fig.~\ref{fig:results}. While the computationally expensive SSA reproduced experimental data relatively accurately, the hybrid approach failed to accurately describe the population average and underestimated the number of bursts.
  Unlike in the CIR internalization model, in both the full SSA and the hybrid method, the nuc/cyt SMAD2 ratio tends to stick to the level before stimulation once the trajectory reached this level again and subsequent stochastic events were less frequently observed. This behaviour was even more prominent in the hybrid model. 
Moreover the full SSA model predicts a higher number of bursts with high amplitude, in comparison to both data and CIR internalization model, which all attain a similar maximum level. Both SSA models overestimate the heights of later bursts, which was not observed in the CIR internalization model.
The predicted bursts thus vary significantly from the ones observed in the experimental data and those predicted by the CIR internalization model.
\url{https://doi.org/10.6084/m9.figshare.19064702}

\paragraph{S8 Table. Comparison to full and hybrid stochastic simulation techniques}
Distance measurements of $\tau$-leaping SSA and a hybrid SSA solver (see S7 Fig.) in comparison to the CIR internalization and the deterministic model to experimental data for a stimulation with 100 pM TGF-β (compare Table~\ref{tab:costs}). Distance in terms of burst statistics of both $\tau$-leaping SSA and hybrid SSA is increased in comparison to the CIR internalization model. \url{https://doi.org/10.6084/m9.figshare.19064621}

\paragraph{S9 Fig. Bursts in the nuc/cyt SMAD2 ratio are caused by anti-correlated changes in nuclear and cytoplasmic SMAD2}
To validate that the bursting events in the nuc/cyt SMAD2 ratio do not arise from technical artefacts, we analyzed the behavior of cytoplasmic as well as nuclear SMAD2 pools in trajectories of single cells. 
If the bursts are related to nuclear translocation, the two pools must be negatively correlated in single cells during bursting events.
Even though over all cells and considered time points, nuclear and cytoplasmic SMAD2 are positively correlated, in the progress of the initial burst, they are indeed negatively correlated. 
In the scatter plot on the left each trajectory is represented by a blue dot and a red dot encoding nuclear and cytoplasmic SMAD2 at the time before and within a burst event respectively. The histogram on the right shows the distribution of angles of the line connecting the initial state to the burst peak in the cell population, clearly indicating that for most cells, the increase in nuclear SMAD2 matched the decrease of the cytoplasmic SMAD2.
This excludes the existence of globally correlated measurement noise and suggests that nuclear and cytoplasmic SMAD2 is negatively correlated within bursts. \url{https://doi.org/10.6084/m9.figshare.19064699}

\paragraph{S10 Fig. Comparison to alternative vesicle model}
To assess the role of receptor sorting into vesicles we considered an alternative model.
This vesicle model assumed the receptors to be attributed to a particular area on the cell surface. We randomly decided whether each vesicle got internalized or not. For each vesicle that gets internalized, all attributed receptors were internalized, independent of their binding states. For simplicity, after internalisation, the receptors are assumed to be released and recycled or degraded according to the ODE model. 
 Thus the underlying hypothesis of vesicles is not yet consistently translated into the new model.
Still, even without parameter fitting, the number of bursting events agreed well to the experimental data. Simulation data and burst analysis is presented analog to Fig.~\ref{fig:results}. While the variance predicted by this model is clearly underestimated, the average height and duration of bursting events were well predicted (see also S11 Table). The variance of the burst properties though, were underestimated. While the results of the vesicle model are promising, further research and modeling are required to better describe the data.
\url{https://doi.org/10.6084/m9.figshare.19064675.v1}

  
\paragraph{S11 Table. Comparison to alternative vesicle model}
Distance measurements of the vesicle model (see S10 Fig.) in comparison to the CIR model and the deterministic model relative to experimental data for stimulation with 100 pM TGF-β (compare Table~\ref{tab:costs}). While the number of bursts predicted by the vesicle model matches the experimental data, the fit in standard deviation and population average is increased compared to the CIR model. \url{https://doi.org/10.6084/m9.figshare.19064660} 

\paragraph{S12 Fig. Ligand degradation}
The free ligand concentration in the CIR internalization model in case of stimulation with 100 pM TGF-β for different population sizes (left) and their difference from a reference ligand concentration computed by assuming population size $n=750$ (right). Temporal fluctuations in the ligand concentration average out for large population sizes. \url{https://doi.org/10.6084/m9.figshare.19064666}
}

\subsection*{Data availability statement}
The source code and data used to produce the results and analyses presented in this manuscript are available from the GitHub repository~\cite{stochSMADGit}.

\subsection*{Acknowledgements}
The experimental data used in this paper was partially produced by Jette Strasen (MDC Berlin). 



\bibliography{stochSMAD}
\appendix
\begin{table}[!ht]
    \caption{Dynamic variables and model equations}\label{tab:model}
	{\scriptsize
          \begin{tabular}{r  p{6cm} @{\hskip 1cm} r p{9cm}}
            \toprule
			 & description & &governing time derivative\\ \midrule
		$L$ & free TGF-β ligand  & $f_L=$ & $E_{1}(t)+ \sum_{i=1}^{N} E_7^{i}\, E_{9}(t)\frac{y_6^{i}}{N\, P_{40}^{i}} - \sum_{i=1}^{N} P_{49}^{i} E_9(t) \frac{L y^i_2 }{N \, P_{40}^{i}} $\\	
		
		$y_1$ & TGF-β receptor I (TGFBR1) on cell surface & $f_1=$ & $E_{2} + E_{6}y_{7}+ P_{15}P_{44}y_{9}+P_{16}P_{52}y_{9}+P_{17}P_{37}y_{3}-P_{37}y_{1}-P_{48}y_{1}y_{6}$\\
		$y_2$ & TGF-β receptor II (TGFBR2) on cell surface & $f_2=$ & $E_{3} + E_{7}E_{9}(t)y_{6} +P_{16}P_{52}y_{9}+P_{18}P_{38}y_{4}-P_{38}y_{2}-P_{49}E_{9}(t)y_{2}L$\\
		$y_3$ & endosomal TGFBR1 & $f_{3}=$ & $ -P_{17}P_{37}y_{3}+P_{36}y_{8}+P_{37}y_{1}-P_{43}y_{3}+P_{44}y_{8} +P_{52}y_{8}$\\
		$y_4$ & endosomal TGFBR2 & $f_{4}=$ & $ -P_{18}P_{38}y_{4}+P_{36}y_{8}+P_{38}y_{2}+P_{43}y_{8}-P_{44}y_{4} +P_{52}y_{8}$\\
		
		$y_5$ & endosomal TGF-β & $f_{5}=$ & $P_{15}P_{44}y_{9} + P_{36}y_{8}+P_{43}y_{8}+P_{44}y_{8}-P_{52}y_{5}$\\
		$y_6$ & complex of TGF-β and TGFBR2 on cell surface &$f_{6}=$ & $ E_{6}y_{7}-E_{7}E_{9}(t)y_{6}-P_{48}y_{1}y_{6}+P_{49}E_{9}(t)y_{2}L$\\
		$y_7$ & activated complex of TGF-β, TGFBR1 and TGFBR2 on cell surface &$f_{7}=$ & $-E_{6}y_{7}+E_{8}y_{9}-P_{14}P_{37}y_{7}+P_{48}y_{1}y_{6}-P_{50}y_{7}y_{17}$\\
		$y_8$ & activated endosomal complex of TGF-β, TGFBR1 and TGFBR2&$f_{8}=$ & $ P_{14}P_{37}y_{7}-P_{36}y_{8}-P_{43}y_{8}-P_{44}y_{8}-P_{52}y_{8}$\\
		$y_{9}$ &  inactivated complex of TGF-β, TGFBR1, TGFBR2 and SMAD7 on cell surface & $f_{9}=$ & $-E_{8}y_{9}-P_{15}P_{44}y_{9} -P_{16}P_{52}y_{9}+P_{50}y_{7}y_{17}$\\
		$y_{10}$ & cytoplasmic phosphorylated SMAD2 & $f_{10}=$ & $ \frac{P_{5}y_{18}}{2}-P_{6}y_{10}+3P_{21}P_{42}y_{14}-3P_{22}P_{51}y_{10}^{3}+P_{41}y_{8}y_{11}+ 2P_{42}y_{13}-P_{45}y_{10}+P_{45}y_{13}+2P_{45}y_{14}+2P_{46}y_{13}-2P_{51}y_{10}^{2}y_{12}$\\
		
		$y_{11}$ & cytoplasmic unphosphorylated SMAD2 & $f_{11}=$ & $ E_{4}+\frac{P_{5}y_{21}}{2}-P_{6}y_{11}-P_{41}y_{8}y_{11}-P_{45}y_{11}$\\
		$y_{12}$ & cytoplasmic SMAD4 & $f_{12}=$ & $ E_{5}+\frac{P_{8}y_{22}}{2}-P_{9}y_{12}+P_{42}y_{13}+P_{45}y_{13}-P_{46}y_{12}-P_{51}y_{10}^{2}y_{12}$\\
		$y_{13}$ & cytoplasmic heterotrimer of two phosphorylated SMAD2 proteins and one SMAD4 protein& $f_{13}=$ & $ -(P_{11} + P_{42} + P_{45} + P_{46})y_{13}+P_{51}y_{10}^{2}y_{12}$\\
		$y_{14}$ & cytoplasmic homotrimer of phosphorylated SMAD2 &$f_{14}=$ & $ -P_{11}y_{14}-P_{21}P_{42}y_{14}+P_{22}P_{51}y_{10}^{3}-P_{45}y_{14}$\\
		$y_{15}$ & cytoplasmic mRNA of SMAD7 & $f_{15}=$ & $\frac{P_{12}y_{16}}{2}-P_{54}y_{15}$\\
		$y_{16}$ & nuclear mRNA of SMAD7 & $f_{16=}$ & $ P_{39}~\frac{y_{19}^{P_{13}}}{P_{2}^{P_{13}}+y_{19}^{P_{13}}} + P_{55}-P_{12}y_{16}-P_{53}y_{16}$\\
		$y_{17}$ & cytoplasmic SMAD7 (feedback protein) & $f_{17}= $ & $E_{8}y_{9} + P_{15}P_{44}y_{9}+P_{16}P_{52}y_{9}+P_{35}y_{15}-P_{47}y_{17}-P_{50}y_{7}y_{17}$\\
		$y_{18}$ & nuclear phosphorylated SMAD2 & $f_{18}= $ & $ -P_{5}y_{18}+2P_{6}y_{10}+3P_{22}P_{41}y_{20}-3P_{21}P_{51}y_{18}^{3}+P_{24}P_{34}y_{19}-P_{34}y_{18}+2P_{42}y_{19} -P_{45}y_{18}+P_{45}y_{19}+2P_{45}y_{20}+2P_{46}y_{19}-2P_{51}y_{18}^{2}y_{22}$\\
		
		$y_{19}$ & nuclear heterotrimer of two phosphorylated SMAD2 proteins and one SMAD4 protein &  $f_{19}= $ & $ 2P_{11}y_{13}-P_{24}P_{34}y_{19}-P_{42}y_{19}-P_{45}y_{19}-P_{46}y_{19}+P_{51}y_{18}^{2}y_{22}$\\
		$y_{20}$ & nuclear homotrimer of phosphorylated SMAD2 & $f_{20}=$ & $ 2P_{11}y_{14}-P_{21}P_{42}y_{20}+P_{22}P_{51}y_{18}^{3}-P_{45}y_{20}$\\
		$y_{21}$ & nuclear unphosphorylated SMAD2& $f_{21}= $ & $ -P_{5}y_{21}+2P_{6}y_{11}+P_{24}P_{34}y_{19}+P_{34}y_{18}-P_{45}y_{21}$\\
            $y_{22}$ & nuclear SMAD4& $f_{22}=$ & $ -P_{8}y_{22}+2P_{9}y_{12}+P_{24}P_{34}y_{19}+P_{42}y_{19}+P_{45}y_{19} -P_{46}y_{22}-P_{51}y_{18}^{2}y_{22}$\\ \bottomrule\\	
\end{tabular}}
\begin{flushleft} 
	Description of the dynamic variables and their corresponding right hand sides used in the model equations \eqref{eq:lig} and \eqref{eq:dyn_con}. For brevity, the dependence of the dynamic concentrations on $i$ (indicating the cell) is omitted in the descriptions of $f_1, \dots, f_{22}$. Details about the kinetic and experimental parameters used here can be found in Tables~\ref{tab:parameters} and~\ref{tab:experiments}.
\end{flushleft}
\end{table}

	\begin{table}
	\caption{Kinetic model parameters}\label{tab:parameters}
	{\scriptsize
	\begin{tabular}{r l l l}\toprule
		 & name in the code& description & block model\\\midrule
$P_{2}$	&	K-mran	&	Smad 7 induction hill parameter  (Michaelis constant)&  \\
$P_{3}$	&	R1-total	&	total number of TGFBR1 receptors&  synthesis \\
$P_{4}$	&	R2-total	&	total number of TGFBR2 receptors& synthesis \\
$P_{5}$	&	S2-export-from-nuc	&	rate of SMAD2 export from the nucleus to the cytoplasm&  \\
$P_{6}$	&	S2-import-to-nuc	&	rate of SMAD2 import from the cytoplasm to the nucleus&  \\
$P_{7}$	&	S2-total	&	total number of SMAD2 proteins& \\
$P_{8}$	&	S4-export-from-nuc	& rate of SMAD4 export from the nucleus to the cytoplasm & \\
$P_{9}$	&	S4-import-to-nuc	& rate of SMAD4 import from the cytoplasm to the nucleus & \\
$P_{10}$	&	S4-total	&	total number of SMAD4 proteins & \\
$P_{11}$	&	Trimer-import-to-nuc	& rate of SMAD trimer import to the nucleus & \\
$P_{12}$	&	export-cytoplasm	& rate of SMAD7 export from the cell & \\
$P_{13}$	&	hill-fact1	&	SMAD7 induction hill parameter  (hill coefficient) & \\
$P_{14}$	&	index-active-Rec-internalize	&	internalization rate of activated receptor complexes &  internalization\\
$P_{15}$	&	index-induced-R2-deg	&	decay rate of SMAD7 inactivated  TGFBR2 receptors  & \\
$P_{16}$	&	index-induced-ligand-deg	&	decay rate of TGF-β bound by inactivated TGFBR2 receptor  & \\
$P_{17}$	&	index-k-out-1-relative-speed	&	speed of the export of TGFBR1 out of the cell relative to its import into the cell & \\
$P_{18}$	&	index-k-out-2-relative-speed	&	speed of the export of TGFBR2 out of the cell relative to its import into the cell & \\
$P_{19}$	&	index-kb-R1	& speed of the unbinding of TGFBR1 from the ligand relative to its binding from the ligand& \\
$P_{20}$	&	index-kb-R2	& speed of the unbinding of TGFBR2 from the ligand relative to its binding from the ligand& \\
$P_{21}$	&	index-kb-homotrimer	& rate of the unbinding of SMAD homotrimers into single SMADs & \\
$P_{22}$ &	index-kf-homotrimer	&	rate of the binding of single SMADs to homotrimers & \\
$P_{23}$	&	index-seq-kb	&	speed of the unbinding of SMAD7 to activated receptor complexes relative to its unbinding& \\
$P_{24}$	&	index-trimer-dephos	&	decay rate of pSMAD within a trimer & \\
$P_{25}$	&	k1	&	dose constant of 1 pM stimulation & \\
$P_{26}$	&	k2	&	dose constant of 2.5 pM stimulation & \\
$P_{27}$	&	k3	&	dose constant of 5 pM stimulation & \\
$P_{28}$	& k4	&	dose constant of 25 pM stimulation & \\
$P_{29}$	&	k5	&	dose constant of 100 pM stimulation & \\
$P_{34}$	&	k-Dephos	&	dephosphorylation rate of SMAD & \\
$P_{35}$	&	k-S7-protein	&	synthesis rate of SMAD7 from mRNA (translation) & \\
$P_{36}$	&	k-disso-Active-Rec	&	dissociation rate of active receptor complexes & end. traffic\\
$P_{37}$	&	k-in-1	&	internalization rate of TGFBR1 (shutteling) & internalization \\
$P_{38}$	&	k-in-2	&	internalization rate of TGFBR2 (shutteling) & internalization\\
$P_{39}$	&	k-induced-S7-production	&	rate of the SMAD signaling induced production of SMAD7 & \\
$P_{40}$	&	k-medium	&	cell volume  & \\
$P_{41}$	&	k-phosphorylation	&	phosphorylation rate of SMAD2 to pSMAD2 & \\
$P_{42}$	&	kb-trimmer	&	dissociation rate of SMAD trimers to SMADs & \\
$P_{43}$	&	kdeg-R1	&	degradation rate of TGFBR1 receptors & degradation \\
$P_{44}$	&	kdeg-R2	&	degradation rate of TGFBR2 receptors & degradation \\
$P_{45}$	&	kdeg-S2	&	degradation rate of SMAD2 proteins& \\
$P_{46}$	&	kdeg-S4	&	degradation rate of SMAD4 proteins& \\
$P_{47}$	&	kdeg-S7	&	degradation rate of SMAD7 proteins& degradation \\
$P_{48}$	&	kf-R1-activation	&	rate of TGFBR1 binding to activated TGFBR2 complexes & receptor/ligand\\
$P_{49}$	&	kf-R2-activation	&	rate of TGFBR2 binding to the ligand& receptor ligand \\
$P_{50}$	&	kf-Seq-S7-Rec	&	 rate of SMAD7 binding to activated receptor complexes & \\
$P_{51}$	&	kf-trimmer	&	rate of SMAD binding to build SMAD trimers & \\
$P_{52}$	&	kin-deg-Ligand	&	degradation rate of TGF-β  & \\
$P_{53}$	&	kmRNA1deg-S7	&	degradation rate of SMAD7 mRNA in the nucleus& \\
 $P_{54}$	&	kmRNAdeg-S7	&	degradation rate of SMAD7 mRNA in the cytoplasm  & \\
$P_{55}$	&	mRNA-prod	&	basal production rate of SMAD7 mRNA (transcription) & synthesis\\ \bottomrule\\
	\end{tabular}
    }
\begin{flushleft} 
	Description of the kinetic parameters used in the model, their name in the code. Indications are given when kinetic parameters are used to introduce noise in a block model. The Parameters $P_{25}$--$P_{29}$ are used for input scaling and are therefore not considered kinetic parameters.
\end{flushleft}
	\end{table}

	\begin{table}
		\caption{Parameters for experimental conditions and substitutions in the model descriptions}\label{tab:experiments}
		{\scriptsize
				\begin{tabular}{r l@{\hskip 1cm} r p{3.5cm}}\toprule 
	$E_{1}(t, \text{1 pM}) =$ &$P_{25} Q^+(0.1, t)\,Q^-(2.1, t)$&
 	$E_{1}(t, \text{2.5 pM}) =$ &$P_{26} Q^+(0.1, t)\,Q^-(2.1, t)$\\
 	$E_{1}(t, \text{5 pM}) =$ &$P_{27} Q^+(0.1, t)\,Q^-(2.1, t)$&
 	$E_{1}(t, \text{25 pM}) =$ &$P_{28} Q^+(0.1, t)\,Q^-(2.1, t)$\\
 	$E_{1}(t, \text{100 pM}) =$ &$P_{29} Q^+(0.1, t)\,Q^-(2.1, t)$&
 	$E_{1}(t, \text{2$\times$2.5 pM}) =$ &$P_{26} Q^+(0.1, t)\,Q^-(2.1, t) + P_{26} Q^+(480.1, t)\,Q^-(482.1, t)$\\
	$E_{2} =$ & $P_{3}\frac{P_{37}~P_{43}^{2}}{P_{17}P_{37}P_{43}+P_{43}^{2}+P_{37}P_{43}}$ &
	$E_{3} =$ & $P_{4}\frac{P_{38}~P_{44}^{2}}{P_{18}P_{38}P_{44}+P_{44}^{2}+P_{38}P_{44}}$\\
	$E_{4} =$ & $P_{7}\frac{P_{6}P_{45}+P_{5}P_{45}+P_{45}^{2}}{P_{5}+P_{45}+2P_{6}}$&
	$E_{5} =$ & $P_{10}\frac{P_{9}P_{46}+P_{8}P_{46}+P_{46}^{2}}{P_{8}+P_{46}+2P_{9}}$ \\
	$E_{6} =$ & $P_{48}P_{19}$&
	$E_{7} =$ & $P_{49}P_{20}$\\
	$E_{8} =$ & $P_{50}P_{23}$&
	$E_{9}(t) =$ & $Q^+(0.1, t)$\\
	$Q^+(\tau, t) =$ & $\frac{1000}{1000 +100\exp(\tau-t)}$&
                                                                 $Q^-(\tau, t) =$ & $\frac{1000}{1000 +100\exp(t - \tau))}$\\ \bottomrule\\
                                  
	\end{tabular}}
\begin{flushleft} 
	 Parameters of experimental conditions that were used in the model (see Table~\ref{tab:model}). The parameter $E_1$ describes the dose dependent bolus induction. The Parameters $E_2$, $E_3$, $E_4$ and $E_5$ describe the production of TGFBR1, TGFBR2, SMAD2 and SMAD4, respectively. Unbinding from complexes is determined by $E_6$ (TGFBR1 from activated complex of TGF-β, TGFBR1 and TGFBR2 ($y_7$)), $E_7$ (TGFBR2 from activated complex of TGF-β and TGFBR2  ($y_6$)) and $E_8$ (SMAD7 from inactivated complex of TGF-β, TGFBR1, TGFBR2 and SMAD7 ($y_9$)). The parameter $E_9$ introduces a short delay in some reactions for numerical stability. 
\end{flushleft}
\end{table}
	
	\begin{table}
		\caption{High order corrections for the path computation scheme}\label{tab:highorder}
		{\scriptsize
				\begin{tabular}{r l}\toprule
				  $G^i_{k,n} = $ & $\sum_{j\in S} \sigma_j \sqrt{p^i_{j,n}}\frac{\partial f_k(t_{n}, L_n, \mathbf{Y}^k_{n}, \mathbf{P}^k_{n}) }{\partial p_j} \left[\Delta Z^i_{j,n} - \frac12 \Delta t \Delta \changed{B}^{i}_{j,n}\right]$ \\
				   $H^{i}_{j,n} = $ & $\frac{\theta_{j} \sigma_{j}(p^{i}_{j,0} - p^{i}_{j,n})}{2\, \sqrt{p^{i}_{j,n}}} \left[\Delta \changed{B}^{i}_{j,n} \Delta t - \Delta Z^{i}_{j,n} \right] -
				   					 \theta_{j}\sigma_{j} \sqrt{p^{i}_{j,n}}\left[ \Delta Z^{i}_{j,n}- \frac12 \Delta \changed{B}^{i}_{j,k}\Delta t \right]$ \\ $\Delta Z^i_{j,n} =$ & $ \frac{1}{2} \sqrt[3]{\Delta t}\left(\mathcal{M}_{j,n}^i + \frac{1}{\sqrt{3}} \mathcal{N}_{j,n}^i\right)$\\ \bottomrule \\
				 \end{tabular}
		}
\begin{flushleft}
	Formulas of the high order corrections used in the numerical scheme. $\mathcal{M}_{j,n}^i$ and $\mathcal{N}_{j,n}^i$ denote independent identically distributed standard normal random variables.
\end{flushleft}
\end{table}
\end{document}